\documentclass[graybox, envcountchap]{svmult}

\usepackage{mathptmx}        
\usepackage{amsmath}
\usepackage{amssymb}
\usepackage{color}
\usepackage{helvet}          
\usepackage{courier}         
\usepackage{dirtree}

\usepackage{makeidx}      
\usepackage{graphicx}        
\usepackage{subfig}

\usepackage{multicol}        
\usepackage[bottom]{footmisc}

\usepackage{hyperref}        
\hypersetup{colorlinks=true,urlcolor=blue}

\usepackage[misc]{ifsym}
\usepackage{tikz}
\usetikzlibrary{positioning, arrows.meta, shapes.geometric, calc}

\makeindex        

\begin{document}
\title{Machine Learning for Event Reconstruction in Imaging Atmospheric Cherenkov Telescopes}
\titlerunning{ML for Event Reconstruction in IACTs}
\author{Antonio Pagliaro and Antonino La Barbera}
\authorrunning{A. Pagliaro and A. La Barbera}
\institute{Antonio Pagliaro (\Letter) and Antonino La Barbera \at INAF IASF Palermo, Via Ugo La Malfa 153, Palermo, I-90146, Italy \\}
%
%
\maketitle
\abstract{Imaging Atmospheric Cherenkov Telescopes (IACTs) have become the leading instruments for exploring the universe at very high energies (VHE), from hundreds of GeV to hundreds of TeV. Together with wide-field observatories based on particle-detector arrays, such as LHAASO, which extend the accessible energy range well into the ultra-high-energy (UHE) domain above 100\,TeV, they form the backbone of modern ground-based gamma-ray astronomy. This chapter provides a review of the critical role machine learning plays in reconstructing the physical properties of particles detected by these instruments. We begin by introducing the fundamental principles of the IACT technique, highlighting the central challenge: distinguishing the rare gamma-ray-initiated air showers from an overwhelming background of cosmic-ray-induced showers. We then detail the standard reconstruction pipeline currently adopted by operating experiments, from image cleaning and morphological parameterization using the Hillas framework to the powerful enhancements offered by stereoscopic observation. Subsequently, we frame the reconstruction problem in the context of modern supervised machine learning, outlining its application to the dual tasks of particle identification (classification) and parameter estimation (regression). The chapter then explores two frontiers of innovation. First, we delve into the use of the temporal dimension of shower images, defining a new lexicon of timing-based features and demonstrating their power to significantly enhance background rejection, particularly at the scientifically crucial low-energy threshold. Second, we investigate the application of advanced ensemble learning techniques, such as gradient boosting and stacking, which move beyond the baseline Random Forest models to achieve superior performance, most notably in mitigating the systematic energy bias. Finally, we synthesize these concepts, discuss the key metrics for performance evaluation, and provide an outlook on the next horizon of IACT analysis, which is increasingly dominated by deep learning paradigms like Convolutional and Graph Neural Networks.}


\section{The Challenge of Ground-Based Gamma-Ray Astronomy}
\label{sec:1}

Gamma-ray astronomy provides a unique window into the most extreme and energetic phenomena in the universe, probing particle acceleration in sources such as supernova remnants, active galactic nuclei, pulsars, and gamma-ray binaries, among others. The Earth's atmosphere is opaque to most of the electromagnetic spectrum; only visible light and a portion of the radio band can reach the ground directly. Gamma rays, in particular those with energies above a few hundred GeV---known as very-high-energy (VHE) gamma rays---are efficiently absorbed high in the atmosphere and cannot be detected directly at ground level. Their study from the ground is made possible by an ingenious indirect detection method: the Imaging Atmospheric Cherenkov Telescope (IACT) technique~\cite{Holder:2012, deNaurois:2015}. It should be noted that IACTs are not the only ground-based technique for VHE gamma-ray detection. Wide-field observatories based on particle-detector arrays---such as LHAASO (Large High Altitude Air Shower Observatory)~\cite{LHAASO:2024}, HAWC, and the future SWGO---detect the secondary particles of Extensive Air Showers (EASs) that reach the ground, rather than their Cherenkov light. These instruments offer a complementary approach: they have a very large field of view and near-continuous duty cycle, making them ideal for surveying the sky and monitoring transient phenomena, and they extend the accessible energy range well into the ultra-high-energy (UHE) domain above 100\,TeV. However, their angular resolution and background rejection capabilities are generally inferior to those of IACTs at overlapping energies, and the machine learning challenges they face---while sharing some conceptual similarities---differ in detail. This chapter focuses on the IACT technique, which currently provides the best sensitivity in the core VHE range from $\sim$20\,GeV to $\sim$100\,TeV.

This section introduces the principles of this technique and outlines the fundamental data analysis challenge that necessitates the use of sophisticated machine learning methods.

\subsection{The Imaging Atmospheric Cherenkov Telescope (IACT) Technique}
\label{subsec:1.1}

When a VHE gamma ray enters the Earth's atmosphere, it interacts with an atomic nucleus, typically producing an electron-positron pair. These secondary particles are still highly energetic and subsequently interact via bremsstrahlung (emitting gamma rays) and further pair production. This cascade of interactions creates a swarm of relativistic secondary particles known as an Extensive Air Shower (EAS)~\cite{deNaurois:2015}. The atmosphere itself acts as a vast calorimeter, converting the energy of the single primary particle into a multitude of lower-energy secondaries~\cite{Holder:2012}.

Many of the charged particles within this shower travel faster than the speed of light in the local medium (the atmosphere). This superluminal motion polarizes the atmospheric molecules, which then emit a faint, coherent cone of bluish-ultraviolet light known as Cherenkov radiation~\cite{Kobzev:2010}. Although Cherenkov photons are emitted throughout the entire development of the shower, which lasts on the order of microseconds, the geometrical properties of the emission cone cause the photons to arrive at the ground nearly simultaneously, producing a brief flash of only a few nanoseconds at any given point within a large "light pool", typically hundreds of meters in diameter. IACTs are essentially large optical reflectors that collect this faint Cherenkov light and focus it onto a camera composed of an array of highly sensitive photomultiplier tubes (PMTs). The result is not a direct image of the astrophysical source, but rather a snapshot---an elliptical ``footprint''---of the Cherenkov light produced by the air shower~\cite{Holder:2012, deNaurois:2015, Scuderi:2022}.

The large size of this light pool is, in fact, the key reason why VHE gamma-ray astronomy is carried out from the ground rather than from space. The astrophysical gamma-ray flux falls steeply with energy, so that above a few hundred GeV it becomes too low for space-based instruments to accumulate statistically significant samples within realistic mission lifetimes: their effective collection area is bounded by the physical size of the detector, typically of order $1~\text{m}^2$. Reaching the sensitivity required at VHE therefore demands collection areas several orders of magnitude larger---of the order of $10^5~\text{m}^2$---which are simply unattainable in orbit. IACTs circumvent this limitation by using the atmosphere itself as an amplifier: the effective collection area of each telescope coincides with the area of the Cherenkov photon pool that reaches the ground, providing a dramatic statistical gain at the cost of the indirect nature of the measurement, which in turn motivates the sophisticated reconstruction techniques discussed in this chapter.

\subsection{From Whipple to CTAO: The Landscape of Cherenkov Observatories}
\label{subsec:1.1b}

The IACT technique was pioneered by the Whipple Observatory in Arizona, which in 1989 achieved the first definitive detection of a VHE gamma-ray source---the Crab Nebula---using a single 10\,m reflector equipped with a camera of 37 PMTs~\cite{Weekes:1989}. This landmark result, enabled by the application of the Hillas image parameterization to reject the hadronic background, established the technique and opened the field of ground-based gamma-ray astronomy.

Building on this success, three major stereoscopic IACT arrays currently dominate the field. \textbf{H.E.S.S.} (High Energy Stereoscopic System), located in Namibia, consists of four 12\,m telescopes and a single large 28\,m telescope, covering energies from $\sim$30\,GeV to $\sim$100\,TeV. H.E.S.S.\ pioneered the use of multivariate analysis techniques, including Boosted Decision Trees (BDTs), for gamma/hadron separation~\cite{Ohm:2009} and has more recently explored deep learning approaches~\cite{Parsons:2020}. \textbf{MAGIC} (Major Atmospheric Gamma Imaging Cherenkov), a pair of 17\,m telescopes on La Palma (Canary Islands), was one of the first experiments to systematically adopt Random Forests for event classification and energy estimation---starting in the late 2000s~\cite{Albert:2008gq}---integrated within its MARS analysis framework. MAGIC was also a pioneer in exploiting timing information for background rejection~\cite{Aliu:2009}. \textbf{VERITAS} (Very Energetic Radiation Imaging Telescope Array System), an array of four 12\,m telescopes in Arizona, similarly employs BDTs and has contributed to refining stereoscopic reconstruction methods~\cite{Holder:2012}.

The next-generation observatory, the \textbf{Cherenkov Telescope Array Observatory (CTAO)}~\cite{CTAConsortium:2019}, represents a transformational leap. CTAO will deploy telescopes of three different sizes---Large-Sized Telescopes (LSTs, 23\,m), Medium-Sized Telescopes (MSTs, $\sim$12\,m), and Small-Sized Telescopes (SSTs, $\sim$4\,m)---across two sites: one in the Northern Hemisphere (La Palma) and one in the Southern Hemisphere (Paranal, Chile). With over 60 telescopes planned, CTAO will achieve a sensitivity approximately one order of magnitude better than current instruments across an energy range from 20\,GeV to 300\,TeV. The \textbf{ASTRI Mini-Array}~\cite{Scuderi:2022, Vercellone:2022}, a precursor and pathfinder for the SST component of CTAO, consists of nine dual-mirror 4\,m telescopes being deployed at the Observatorio del Teide in Tenerife. The ASTRI Mini-Array has a dual role: on the one hand, it is a fully fledged scientific observatory designed to carry out an independent astrophysical programme; on the other, it serves as a technological and methodological pathfinder for the development of novel analysis techniques for future large-scale arrays, and it is in this context that the machine learning innovations discussed in this chapter have been developed and validated.

\subsection{Electromagnetic vs. Hadronic Showers: The Signal and the Noise}
\label{subsec:1.2}

The IACT technique is not exclusively sensitive to gamma rays. The Earth is constantly bombarded by a much higher flux of charged cosmic rays---primarily protons and heavier atomic nuclei, but also electrons and positrons. Cosmic-ray electrons and positrons initiate purely electromagnetic cascades that are virtually indistinguishable from gamma-ray-induced showers on an event-by-event basis. However, because the cosmic-ray lepton flux is nearly isotropic whereas gamma rays from astrophysical sources arrive from well-defined directions, this irreducible background is handled statistically through background-subtraction techniques (see Section~\ref{subsec:6.3}) rather than at the single-event level. The dominant background, however, consists of hadronic cosmic rays (protons and nuclei), whose shower development is governed by hadronic interactions that are fundamentally different from the purely electromagnetic processes of a gamma-ray shower.

\begin{itemize}
    \item \textbf{Signal (Gamma-rays):} A gamma-ray-initiated shower is a clean, purely electromagnetic cascade. Its development is relatively smooth and predictable, resulting in a Cherenkov image that is typically compact, regular, and well-described by an ellipse~\cite{Holder:2012, deNaurois:2015}.
    \item \textbf{Background (Cosmic Rays):} A cosmic-ray-initiated (hadronic) shower is a more complex and chaotic process. The initial hadronic interactions produce a rich variety of secondary particles---pions, kaons, nucleon fragments, and other hadrons~\cite{Heck:1998}. Neutral pions decay into gamma rays, initiating electromagnetic sub-showers, while charged pions can decay into muons. These muons are highly penetrating and can travel far from the primary shower axis. This complex development leads to Cherenkov images that are typically more irregular, patchy, and spatially extended than their gamma-ray counterparts~\cite{Holder:2012, deNaurois:2015}.
\end{itemize}

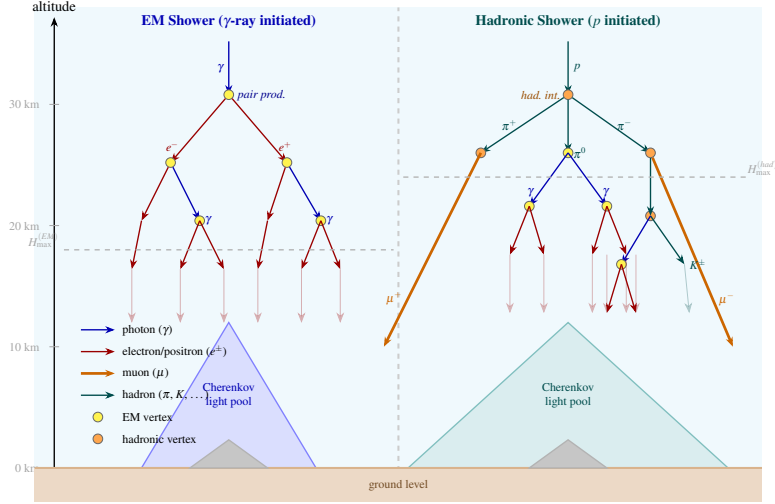
\begin{figure}[t]
\centering
\resizebox{0.88\textwidth}{!}{%
\begin{tikzpicture}[
    photon/.style={-{Stealth[length=2mm]}, thick, blue!70!black},
    lepton/.style={-{Stealth[length=2mm]}, thick, red!60!black},
    muon/.style={-{Stealth[length=2mm]}, ultra thick, orange!80!black},
    pion/.style={-{Stealth[length=2mm]}, thick, teal!60!black},
    vtx/.style={circle, fill=yellow!80, draw=black!60, thin, inner sep=2pt},
    hvtx/.style={circle, fill=orange!70, draw=black!60, thin, inner sep=2pt},
    every node/.style={font=\small}
]

\fill[cyan!5] (-7.5,0) rectangle (7.5,9.5);

\draw[gray!40, very thick, dashed] (0,-0.7) -- (0,9.5);

\draw[-{Stealth[length=2mm]}, thick] (-7.1,0) -- (-7.1,9.3)
    node[above, font=\small] {altitude};
\foreach \y/\lab in {0/{0 km}, 2.5/{10 km}, 5.0/{20 km}, 7.5/{30 km}} {
    \draw[gray!70, thin] (-7.2,\y) -- (-7.0,\y);
    \node[left, font=\scriptsize, gray!70] at (-7.3,\y) {\lab};
}

\node[blue!70!black, font=\small\bfseries, align=center] at (-3.5,9.2)
    {EM Shower ($\gamma$-ray initiated)};
\draw[photon] (-3.5,8.8) -- (-3.5,7.7)
    node[left, midway, font=\scriptsize] {$\gamma$};
\node[vtx] (v0) at (-3.5,7.7) {};
\node[right=0.08cm, font=\scriptsize, blue!55!black] at (v0)
    {\textit{pair prod.}};
\draw[lepton] (v0) -- (-4.7,6.3)
    node[left, near end, font=\scriptsize] {$e^-$};
\draw[lepton] (v0) -- (-2.3,6.3)
    node[right, near end, font=\scriptsize] {$e^+$};
\node[vtx] (v1) at (-4.7,6.3) {};
\draw[lepton] (v1) -- (-5.3,5.1);
\draw[photon] (v1) -- (-4.1,5.1)
    node[right, font=\scriptsize] {$\gamma$};
\node[vtx] (v2) at (-2.3,6.3) {};
\draw[lepton] (v2) -- (-2.7,5.1);
\draw[photon] (v2) -- (-1.6,5.1)
    node[right, font=\scriptsize] {$\gamma$};
\node[vtx] at (-4.1,5.1) {};
\draw[lepton] (-4.1,5.1) -- (-4.5,4.1);
\draw[lepton] (-4.1,5.1) -- (-3.6,4.1);
\node[vtx] at (-1.6,5.1) {};
\draw[lepton] (-1.6,5.1) -- (-2.0,4.1);
\draw[lepton] (-1.6,5.1) -- (-1.2,4.1);
\draw[lepton] (-5.3,5.1) -- (-5.5,4.1);
\draw[lepton] (-2.7,5.1) -- (-2.9,4.1);
\draw[gray!60, dashed, thick] (-6.9,4.5) -- (-0.1,4.5);
\node[gray!70, font=\scriptsize, left] at (-6.9,4.7) {$H_{\max}^{(EM)}$};
\foreach \x in {-5.5,-4.5,-3.6,-2.9,-2.0,-1.2} {
    \draw[lepton, opacity=0.30, thin] (\x,4.1) -- (\x,3.0);
}
\fill[blue!20, opacity=0.60] (-5.3,0) -- (-3.5,3.0) -- (-1.7,0) -- cycle;
\draw[blue!50, thick] (-5.3,0) -- (-3.5,3.0) -- (-1.7,0);
\node[blue!70!black, font=\scriptsize, align=center] at (-3.5,1.5)
    {Cherenkov\\light pool};
\node[font=\scriptsize, gray!70] at (-3.5,-0.4)
    {\textit{compact and regular}};

\node[teal!60!black, font=\small\bfseries, align=center] at (3.5,9.2)
    {Hadronic Shower ($p$ initiated)};
\draw[pion] (3.5,8.8) -- (3.5,7.7)
    node[right, midway, font=\scriptsize] {$p$};
\node[hvtx] (h0) at (3.5,7.7) {};
\node[left=0.05cm, font=\scriptsize, orange!60!black] at (h0)
    {\textit{had. int.}};
\draw[pion] (h0) -- (1.7,6.5)
    node[left, midway, font=\scriptsize] {$\pi^+$};
\draw[pion] (h0) -- (3.5,6.5)
    node[right, font=\scriptsize] {$\pi^0$};
\draw[pion] (h0) -- (5.2,6.5)
    node[right, midway, font=\scriptsize] {$\pi^-$};
\node[vtx] at (3.5,6.5) {};
\draw[photon] (3.5,6.5) -- (2.7,5.4)
    node[left, near end, font=\scriptsize] {$\gamma$};
\draw[photon] (3.5,6.5) -- (4.3,5.4)
    node[right, near end, font=\scriptsize] {$\gamma$};
\node[vtx] at (2.7,5.4) {};
\draw[lepton] (2.7,5.4) -- (2.3,4.4);
\draw[lepton] (2.7,5.4) -- (3.0,4.4);
\node[vtx] at (4.3,5.4) {};
\draw[lepton] (4.3,5.4) -- (4.0,4.4);
\draw[lepton] (4.3,5.4) -- (4.7,4.4);
\node[hvtx] at (1.7,6.5) {};
\draw[muon] (1.7,6.5) -- (-0.3,2.5)
    node[left, near end, font=\scriptsize, orange!80!black] {$\mu^+$};
\node[hvtx] (pim) at (5.2,6.5) {};
\draw[muon] (pim) -- (6.9,2.5)
    node[right, near end, font=\scriptsize, orange!80!black] {$\mu^-$};
\node[hvtx] at (5.2,5.2) {};
\draw[pion] (pim) -- (5.2,5.2);
\draw[pion] (5.2,5.2) -- (5.9,4.2)
    node[right, font=\scriptsize] {$K^\pm$};
\draw[photon] (5.2,5.2) -- (4.6,4.2);
\node[vtx] at (4.6,4.2) {};
\draw[lepton] (4.6,4.2) -- (4.3,3.2);
\draw[lepton] (4.6,4.2) -- (4.9,3.2);
\draw[gray!60, dashed, thick] (0.1,6.0) -- (7.2,6.0);
\node[gray!70, font=\scriptsize, right] at (7.1,6.2) {$H_{\max}^{(had)}$};
\foreach \x in {2.3,3.0,4.0,4.7,4.3,4.9} {
    \draw[lepton, opacity=0.28, thin] (\x,4.4) -- (\x,3.2);
}
\draw[pion, opacity=0.28, thin] (5.9,4.2) -- (6.0,3.2);
\fill[teal!20, opacity=0.60] (0.2,0) -- (3.5,3.0) -- (6.8,0) -- cycle;
\draw[teal!50, thick] (0.2,0) -- (3.5,3.0) -- (6.8,0);
\node[teal!60!black, font=\scriptsize, align=center] at (3.5,1.5)
    {Cherenkov\\light pool};
\node[font=\scriptsize, gray!70] at (3.5,-0.4)
    {\textit{wider and irregular}};

\fill[brown!30] (-7.5,-0.7) rectangle (7.5,0);
\draw[brown!60, very thick] (-7.5,0) -- (7.5,0);
\node[font=\scriptsize, brown!60!black] at (0,-0.35) {ground level};

\draw[fill=gray!45, draw=gray!60, thick]
    (-4.3,0) -- (-3.5,0.58) -- (-2.7,0) -- cycle;
\draw[fill=gray!45, draw=gray!60, thick]
    (2.7,0) -- (3.5,0.58) -- (4.3,0) -- cycle;

\draw[photon]  (-6.55,2.85) -- (-5.85,2.85);
\node[right, font=\scriptsize] at (-5.80,2.85) {photon ($\gamma$)};
\draw[lepton]  (-6.55,2.40) -- (-5.85,2.40);
\node[right, font=\scriptsize] at (-5.80,2.40) {electron/positron ($e^\pm$)};
\draw[muon]    (-6.55,1.95) -- (-5.85,1.95);
\node[right, font=\scriptsize] at (-5.80,1.95) {muon ($\mu$)};
\draw[pion]    (-6.55,1.50) -- (-5.85,1.50);
\node[right, font=\scriptsize] at (-5.80,1.50) {hadron ($\pi$, $K$, \ldots)};
\node[vtx]  at (-6.20,1.05) {};
\node[right, font=\scriptsize] at (-5.80,1.05) {EM vertex};
\node[hvtx] at (-6.20,0.60) {};
\node[right, font=\scriptsize] at (-5.80,0.60) {hadronic vertex};

\end{tikzpicture}%
}
\caption{Schematic development of an extensive air shower for a primary gamma ray
(left) and a primary cosmic-ray proton (right). A gamma-ray-induced shower develops
exclusively through electromagnetic processes: successive pair production
($\gamma \to e^+e^-$) and bremsstrahlung ($e^\pm \to e^\pm\gamma$) create a narrow,
geometrically regular cascade whose shower maximum $H_{\max}^{(EM)}$ typically occurs
at an altitude of $\sim$8--12\,km. The resulting Cherenkov light pool at ground level
is compact and regular, with a typical diameter of $\sim$100--250\,m. A hadronic
shower involves additional hadronic interactions that produce pions and kaons: neutral
pions ($\pi^0$) decay into photon pairs that seed electromagnetic sub-showers, while
charged pions ($\pi^\pm$) decay into penetrating muons ($\mu$) that escape the shower
core at large angles, spreading the shower laterally. The resulting Cherenkov light
pool is wider and spatially irregular, and the shower maximum $H_{\max}^{(had)}$
typically occurs higher in the atmosphere than for an electromagnetic shower of
equivalent energy---because hadronic interactions have larger cross-sections and
deposit energy earlier in the atmosphere. These morphological differences in the
Cherenkov images recorded by the IACT cameras are the basis for gamma/hadron
separation.}
\label{fig:eas_comparison}
\end{figure}

The key morphological differences between electromagnetic and hadronic showers are illustrated schematically in Fig.~\ref{fig:eas_comparison}.

The core challenge of IACT data analysis stems from the vast disparity in the flux of these particles. For every VHE gamma ray detected from even the brightest sources in the sky, such as the Crab Nebula, the telescopes are triggered by hundreds to thousands of background cosmic-ray events in the same energy range~\cite{Pagliaro:2023, Ohm:2009}. This profound signal-to-background imbalance means that the success of any IACT observation hinges on the ability to perform highly efficient background rejection.

\subsection{From Cherenkov Photons to Digital Images: The Data Reconstruction Problem}
\label{subsec:1.3}

The ultimate goal of IACT data analysis is to infer the properties of the primary particle from the digital images recorded by the telescope cameras. This process, known as event reconstruction, can be broken down into two primary machine learning tasks:

\begin{enumerate}
    \item \textbf{Particle Identification (Classification):} The task of distinguishing the rare gamma-ray signal events from the overwhelming hadronic background. This is commonly referred to as "gamma/hadron separation."
    \item \textbf{Parameter Estimation (Regression):} The task of reconstructing the physical properties of the primary particle, namely its energy and arrival direction in the sky, from the shower image(s).
\end{enumerate}

At first glance, this appears to be a simple problem of sorting regular elliptical shapes from irregular ones. However, the development of air showers is an inherently stochastic process. Some gamma-ray showers can produce less regular images, while some hadronic showers can, by chance, produce images that closely mimic a typical gamma-ray event. Furthermore, as noted above, cosmic-ray electrons and positrons produce electromagnetic showers that are intrinsically identical to those initiated by gamma rays, adding an additional layer of confusion. The challenge is therefore not one of simple shape identification but of recognizing subtle patterns within a high-dimensional feature space defined by the intensities, spatial locations, and arrival times of photons across hundreds or thousands of camera pixels. The classes of "gamma-like" and "hadron-like" events have significant overlap in this feature space. This inherent complexity, coupled with the need to operate in a low signal-to-noise environment at the photon-counting level, makes simple selection criteria insufficient. It is precisely this challenge that makes machine learning---which excels at identifying complex, non-linear patterns in noisy, high-dimensional data---an indispensable tool for modern Cherenkov astronomy.

\begin{figure}
\centering
  \begin{tabular}{cc}
    \includegraphics[width=5cm,height=5cm]{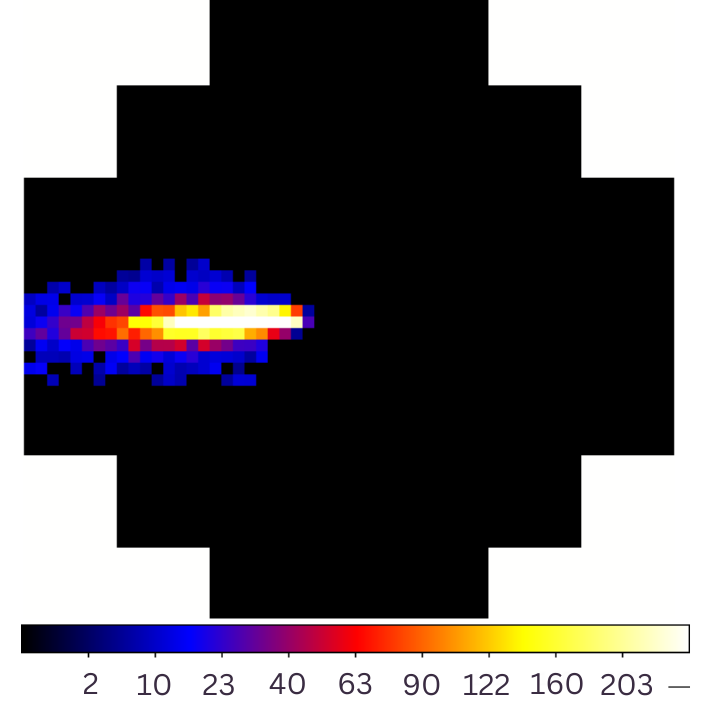} & \includegraphics[width=5cm,height=5cm]{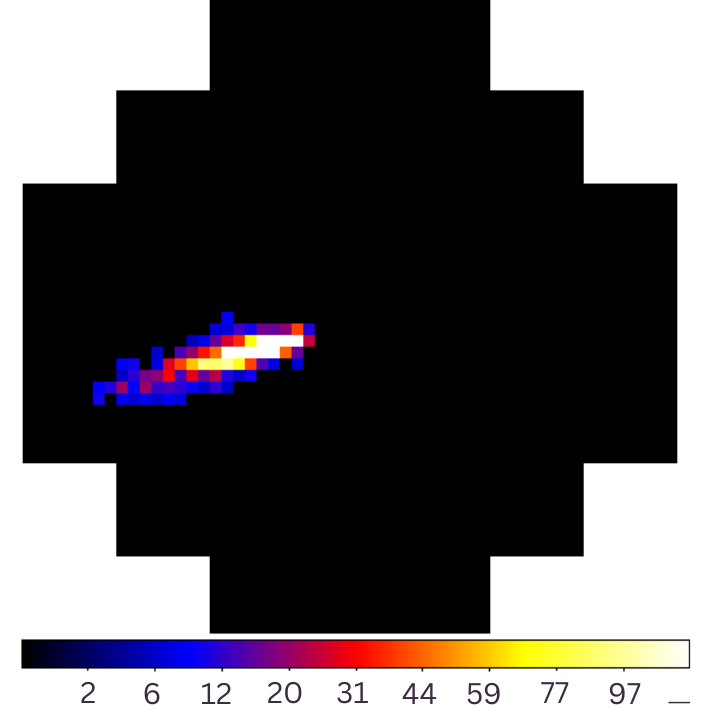} \\ \includegraphics[width=5cm,height=5cm]{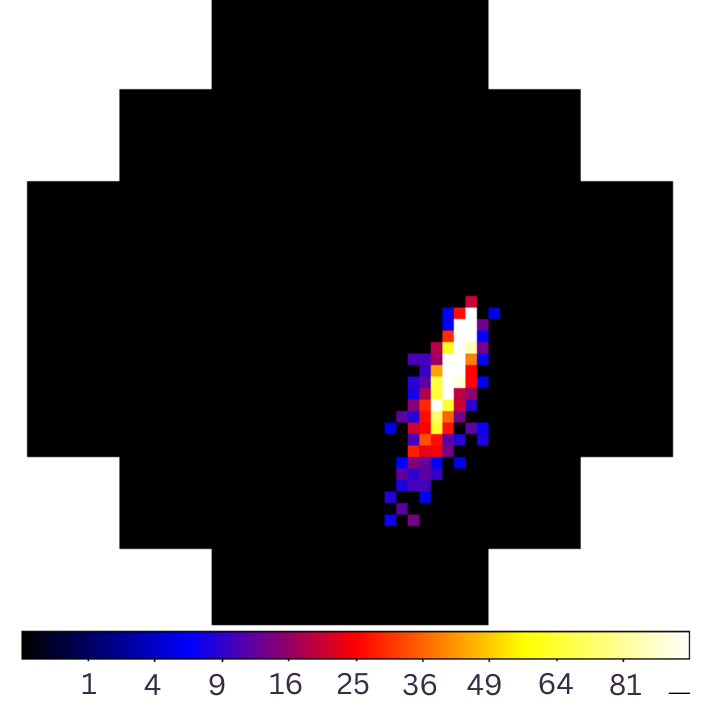} &
    \includegraphics[width=5cm,height=5cm]{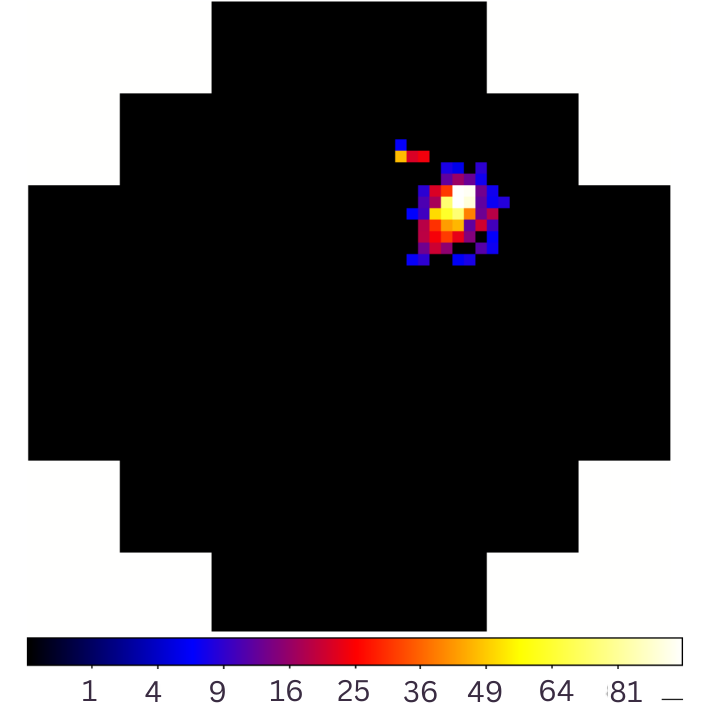} \\
  \end{tabular}
\caption{Cherenkov shower images simulated as observed by the ASTRI MiniArray telescopes, illustrating the morphological differences between different types of events (the contour matches that of the ASTRI telescope camera). The~images include two gamma ray events (\textbf{Top}), a~hadronic event with similarities to gammas (\textbf{Bottom Right}), and~a distinct hadronic event (\textbf{Bottom Left}). These images showcase the often faint but discernible variations in the Cherenkov shower~morphology. Reprinted from \cite{Pagliaro:2023} licensed under CC BY 4.0.\label{fig:astri_shower_images}}
\end{figure}

\section{The Standard Reconstruction Pipeline: From Pixels to Physics}
\label{sec:2}

Before the widespread adoption of end-to-end deep learning, IACT data analysis followed a multi-stage pipeline designed to reduce the raw, high-dimensional camera data into a small set of physically and geometrically meaningful parameters. These parameters then serve as the input features for machine learning models. This section outlines the key steps of this standard reconstruction pipeline.

\subsection{Image Cleaning and Calibration}
\label{subsec:2.1}

The raw data from an IACT camera consists of a time-series of signal amplitudes for each pixel (a short "movie" of the Cherenkov flash). The first step is to calibrate these signals, converting them into a standardized unit, typically photoelectrons (p.e.), which is proportional to the number of Cherenkov photons detected~\cite{LaParola:2025}.

Following calibration, the image must be "cleaned." The camera is constantly exposed to the diffuse light of the night sky (Night Sky Background, or NSB), as well as electronic noise. Image cleaning algorithms are designed to remove these noise contributions and identify only those pixels that are likely part of the true Cherenkov shower image. A common method involves a two-level thresholding procedure~\cite{Hillas:1985}:
\begin{enumerate}
    \item A pixel is selected as a ``core'' pixel if its signal exceeds a higher threshold, called the \emph{picture} threshold.
    \item A pixel adjacent to a core pixel is retained as a ``boundary'' pixel if its signal exceeds a lower threshold, called the \emph{boundary} threshold.
\end{enumerate}
All other pixels are discarded as noise. This process is critical for ensuring that subsequent parameterization is performed on a noise-suppressed representation of the shower.

\subsection{Morphological Parameterization: The Hillas Framework}
\label{subsec:2.2}

A pivotal development in IACT analysis was the introduction of a method to parameterize the cleaned shower image by calculating the first and second moments of its light distribution~\cite{Hillas:1985}. This technique was pioneered by A.~Michael Hillas (1932--2017), a British cosmic-ray physicist at the University of Leeds who played a foundational role in ground-based gamma-ray astronomy. At the 19th International Cosmic Ray Conference in 1985, Hillas proposed characterizing the Cherenkov images through a set of simple geometric parameters derived from the moments of the light distribution, an approach that proved decisive for the first detection of a gamma-ray source by the Whipple telescope in 1989~\cite{Weekes:1989}. His parameterization reduces the information from hundreds of pixel values to a handful of descriptive parameters that capture the image's essential geometric and photometric properties~\cite{Hillas:1985, Pagliaro:2023}. This framework remains the foundation of most modern IACT analysis. In his original work~\cite{Hillas:1985}, Hillas introduced parameters such as \texttt{Width}, \texttt{Length}, \texttt{Miss}, \texttt{Azwidth}, \texttt{Distance}, and \texttt{Alpha}. Over the years, the community has refined and extended this set. The key parameters in common use today include:

\begin{itemize}
    \item \textbf{Size:} The total integrated charge (in p.e.) in the cleaned image, which is correlated with the primary particle's energy.
    \item \textbf{Length and Width:} The major and minor axes of the ellipse that best fits the light distribution in the image. The elongation and width of the image depend on the viewing geometry: showers arriving along the telescope axis appear nearly circular, while off-axis showers produce elongated ellipses. For a given geometry, gamma-ray images tend to be narrower than hadronic images owing to the more compact lateral development of electromagnetic cascades.
    \item \textbf{Dist:} The angular distance from the center of the image (its centroid) to the center of the camera's field of view.
    \item \textbf{Alpha:} The angle between the major axis of the image ellipse and the line connecting the image centroid to the camera center. For gamma rays from a point source at the center of the field of view, this angle is expected to be small.
    \item \textbf{Conc:} The concentration of light in the image, defined as the ratio of the charge in the two brightest pixels to the total image \texttt{Size}. Gamma-ray images tend to be more concentrated.
\end{itemize}

\begin{figure}[t]
\centering
\begin{tikzpicture}[
    every node/.style={font=\small},
    >=Stealth,
    dim/.style={<->, thick}
]

\draw[thick, gray!55, fill=gray!7] (0,0) circle (4.5cm);
\node[gray!60, font=\scriptsize] at (0,4.15) {camera field of view};

\begin{scope}[rotate around={38:(1.0,0.4)}]
    \fill[blue!20, opacity=0.75] (1.0,0.4) ellipse (2.7cm and 0.9cm);
    \draw[blue!55, thick]        (1.0,0.4) ellipse (2.7cm and 0.9cm);
\end{scope}

\foreach \px/\py in {
    -0.15/-0.45, 0.15/-0.25, 0.40/-0.05, 0.65/0.10,
    0.90/0.30, 1.15/0.45, 1.40/0.60, 1.65/0.80,
    1.90/0.95, 2.15/1.15, 2.40/1.35,
    0.50/0.25, 0.85/0.60, 1.25/0.20, 1.55/0.85,
    1.80/0.65, 2.05/1.25} {
    \draw[blue!75!black, opacity=0.35, very thin]
        (\px-0.13,\py-0.13) rectangle (\px+0.13,\py+0.13);
}

\draw[black, very thick] (-0.15,-0.15) -- (0.15,0.15);
\draw[black, very thick] (-0.15, 0.15) -- (0.15,-0.15);
\node[below left, font=\scriptsize, align=right] at (-0.18,-0.18)
    {camera center\\(source position)};

\fill[red!80!black] (1.0,0.4) circle (3.5pt);
\node[font=\scriptsize, red!80!black, anchor=north west]
    at (1.18,0.32) {centroid $C$};

\draw[dim, blue!70!black]
    (-1.13,-1.26) -- (3.13,2.06);
\node[font=\scriptsize\bfseries, blue!70!black, rotate=38, anchor=south]
    at (2.40,1.55) {\textbf{LENGTH}};

\draw[dim, blue!50!black]
    (0.45,1.11) -- (1.55,-0.31);
\node[font=\scriptsize\bfseries, blue!50!black, anchor=north west]
    at (1.58,-0.28) {\textbf{WIDTH}};

\draw[->, thick, red!65!black, dashed]
    (0.06,0.024) -- (0.94,0.376);
\node[below, font=\scriptsize\bfseries, red!65!black, yshift=-2pt]
    at (0.45,0.12) {\textbf{DIST}};

\draw[red!65!black, dashed, thin]
    (1.0,0.4) -- (1.75,0.70);

\draw[green!50!black, thick]
    ({1.0+0.85*cos(21.8)},{0.4+0.85*sin(21.8)})
    arc[start angle=21.8, end angle=38, radius=0.85cm];
\node[font=\scriptsize\bfseries, green!50!black]
    at ({1.0+1.20*cos(29.9)},{0.4+1.20*sin(29.9)})
    {$\boldsymbol{\alpha}$ (\textbf{Alpha})};

\draw[->, gray!55, thin, bend right=18]
    (-2.7,2.6) to (0.60,0.60);
\node[align=center, font=\scriptsize, gray!65] at (-3.05,3.0)
    {\textbf{SIZE}\\(total charge, p.e.)};

\node[draw=gray!50, rounded corners=3pt, fill=white,
      inner sep=5pt, font=\scriptsize, align=left,
      text width=3.3cm]
    at (-2.8,-2.8) {
    \textbf{Parameter summary}\\[3pt]
    \textcolor{blue!70!black}{\textbf{Length}} \,major axis\\
    \textcolor{blue!50!black}{\textbf{Width}}  \;\,minor axis\\
    \textcolor{red!65!black}{\textbf{Dist}}    \;\;\,centroid--center\\
    \textcolor{green!50!black}{\textbf{Alpha}} \;\;axis--Dist angle\\[2pt]
    \textit{Size} = total charge
};

\end{tikzpicture}
\caption{Illustration of the standard Hillas parameterization applied to a
Cherenkov shower image (blue ellipse) in an IACT camera. The image represents
the cleaned pixel charge distribution, schematically rendered as a grid of
pixels. \textbf{Length} and \textbf{Width} are the major and minor semi-axes of
the fitted ellipse, characterizing the longitudinal and lateral extent of the
shower; gamma-ray images tend to be narrower (smaller \texttt{Width}) owing to
the more collimated development of electromagnetic cascades. \textbf{Dist} is
the angular distance from the camera center---which corresponds to the nominal
source direction in an on-axis observation---to the image centroid $C$.
\textbf{Alpha} ($\alpha$) is the angle at the centroid between the major axis
and the line connecting the centroid to the camera center; for gamma rays from
a point source, \texttt{Alpha} is expected to be small, as the image ellipse
points approximately toward the source. \textbf{Size} (indicated by the arrow)
is the total integrated charge of the cleaned image in photoelectrons, which
serves as the primary proxy for the energy of the primary particle.}
\label{fig:hillas_params}
\end{figure}
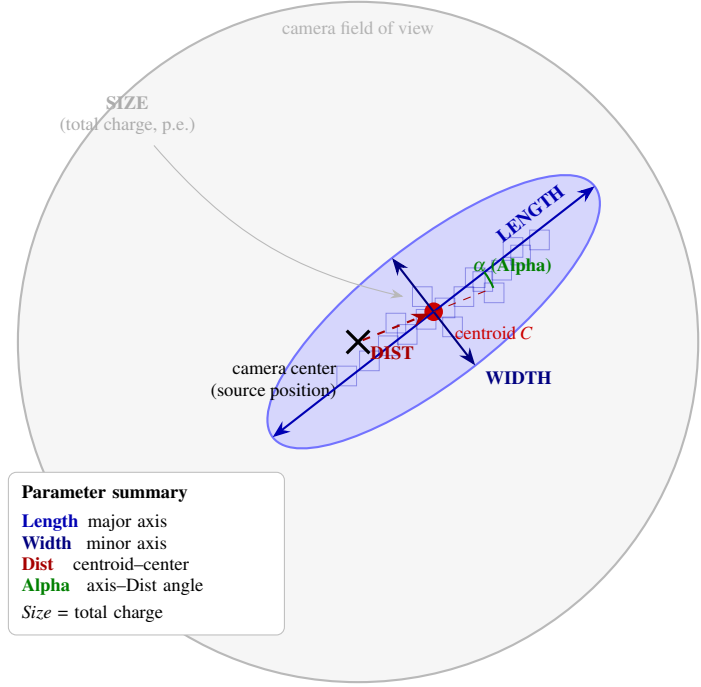

The geometric meaning of these parameters is illustrated in Fig.~\ref{fig:hillas_params}.

This parameterization transforms the abstract problem of image classification into a more tractable problem of classification in a low-dimensional feature space.

\subsection{Stereoscopic Reconstruction: Combining Multiple Viewpoints}
\label{subsec:2.3}

A single telescope provides a 2D projection of the 3D air shower. The major axis of the ellipse that models the Cherenkov image defines a line on the focal plane along which the arrival direction of the primary particle must lie, but a single image cannot resolve the exact position along this line nor the head-tail ambiguity inherent in the elliptical fit. The use of an array of multiple telescopes observing the same shower simultaneously---a technique known as stereoscopy---resolves these ambiguities: the intersection of the major axes from different telescope images yields an estimate of the arrival direction, dramatically improving reconstruction performance~\cite{Holder:2012, deNaurois:2015}.

When two or more telescopes view the same shower, the major axes of their respective image ellipses can be projected onto the sky. The intersection of these axes provides a highly accurate, geometric reconstruction of the shower's arrival direction~\cite{Pagliaro:2023}. This technique also allows for the reconstruction of the shower's core location on the ground (the impact point) and the height in the atmosphere at which the shower reached its maximum development ($H_{max}$). The shower maximum is the depth at which the number of secondary particles in the cascade reaches its peak; beyond this point, energy losses dominate and the number of particles decreases. For electromagnetic showers, the depth of maximum scales logarithmically with the primary energy, while hadronic showers of the same energy typically reach their maximum higher in the atmosphere due to the larger interaction cross-sections involved~\cite{Heck:1998}. These stereoscopic parameters, such as the reconstructed impact distance for each telescope and $H_{max}$, provide powerful additional information that significantly enhances both energy reconstruction and gamma/hadron separation~\cite{Pagliaro:2023}, paralleling the approach adopted in H.E.S.S.~\cite{Ohm:2009} and MAGIC~\cite{Albert:2008gq}. The number of telescopes that participated in the event reconstruction (\texttt{NusedTel}) is itself a valuable parameter.

\section{A Modern Approach: Framing Reconstruction as a Machine Learning Task}
\label{sec:3}

The standard pipeline provides a set of powerful, physically motivated features. The modern approach to IACT analysis uses these features as input to supervised machine learning algorithms~\cite{Bishop:2006, Hastie:2009}, which are trained to perform the classification and regression tasks defined in Section \ref{sec:1}.

\subsection{Supervised Learning for IACTs: Classification and Regression}
\label{subsec:3.1}

Before describing the specific algorithms used in IACT analysis, it is useful to briefly introduce the core concepts of supervised machine learning for readers less familiar with this field. In supervised learning, an algorithm is presented with a set of training examples for which the correct answer is already known. From these examples, the algorithm learns a general rule---a mathematical function---that maps input variables (called \emph{features}) to the desired output. Once trained, this function can be applied to new, unseen data to make predictions. Supervised learning problems fall into two broad categories~\cite{Bishop:2006, Hastie:2009}:
\begin{itemize}
    \item \textbf{Classification:} the output is a discrete category or label (e.g., ``gamma ray'' or ``hadron''). The algorithm learns to assign each new event to one of a finite set of classes.
    \item \textbf{Regression:} the output is a continuous numerical value (e.g., the energy of the primary particle in TeV). The algorithm learns to predict a quantity from the input features.
\end{itemize}

Supervised learning requires a large dataset of labeled examples for training. In the context of IACTs, this ``ground truth'' data is generated through extensive and computationally expensive Monte Carlo (MC) simulations~\cite{Holder:2012, deNaurois:2015}. Software packages like CORSIKA are used to simulate the development of air showers from primary particles of known type, energy, and direction~\cite{Heck:1998}. The resulting Cherenkov light pool is then passed through a detailed simulation of the telescope array's response (e.g., using \texttt{sim\_telarray}), producing realistic digital images~\cite{Bernlohr:2008}.

This process yields vast libraries of simulated events for which the true primary particle properties are known. These libraries form the training, validation, and testing sets for the machine learning models.

\begin{itemize}
    \item \textbf{Gamma/Hadron Separation as Classification:} The task is to train a classifier on the MC-generated features. The model learns the decision boundaries in the multi-dimensional feature space that best separate gamma-ray-like events from hadron-like events. In this context, a decision boundary is the hypersurface in the feature space along which the classifier's output transitions from one class to the other; it is ``non-linear'' when it cannot be described by a simple hyperplane, as is typically the case for the complex distributions encountered in IACT data~\cite{Bishop:2006}. When applied to new data, the classifier outputs a score, typically between 0 and 1, known as \texttt{gammaness}. A high \texttt{gammaness} value indicates that the event is more consistent with a gamma-ray origin; however, it should be noted that the \texttt{gammaness} score is not, in general, a calibrated probability but rather a discriminant variable whose absolute value depends on the training sample composition and the algorithm used~\cite{Pagliaro:2023}.
    \item \textbf{Parameter Estimation as Regression:} Similarly, regression models are trained on the MC features to predict the primary particle's energy and arrival direction. For energy reconstruction, the model learns the intricate relationship between image properties (like \texttt{Size} and impact distance) and the true energy of the primary particle~\cite{Pagliaro:2023}. For direction reconstruction, the geometric intersection of image axes in stereoscopic mode (Section~\ref{subsec:2.3}) already provides a robust estimate; however, machine learning can refine this estimate by learning systematic corrections that depend on the event topology, such as the number of triggered telescopes and the distance of the shower core from the array center. In single-telescope mode, direction reconstruction is considerably more challenging, as the position of the source along the major axis of the image is ambiguous (the ``head-tail'' degeneracy); ML methods trained on asymmetry-related features, such as the skewness of the light distribution, can partially resolve this ambiguity~\cite{Holder:2012}.
\end{itemize}

\subsection{Feature Engineering I: Morphological and Stereoscopic Parameters}
\label{subsec:3.2}

The performance of any machine learning model is fundamentally dependent on the quality of its input features. The classical reconstruction pipeline described in Section \ref{sec:2} is, in essence, a sophisticated feature engineering process. Table \ref{tab:morph_params} summarizes the key morphological and stereoscopic parameters commonly used as input features for machine learning models in IACT analysis, drawing from the parameters used in the analysis for the ASTRI Mini-Array~\cite{Pagliaro:2023}; equivalent feature sets underpin the standard analyses of MAGIC~\cite{Albert:2008gq} and H.E.S.S.~\cite{Ohm:2009}.

\begin{table}
\caption{Key Morphological and Stereoscopic Parameters for IACT Event Reconstruction. These parameters, derived from the standard Hillas framework and stereoscopic analysis, form the standard feature set for machine learning models.}
\label{tab:morph_params}       
\begin{tabular}{p{2.5cm}p{9cm}}
\hline\noalign{\smallskip}
Parameter & Physical Interpretation  \\
\noalign{\smallskip}\svhline\noalign{\smallskip}
$log_{10}(SIZE)$ & The logarithm of the total charge in the cleaned image. A primary proxy for the event's energy. \\
WIDTH & The minor axis of the best-fit image ellipse. Measures the lateral spread of the shower image. \\
LENGTH & The major axis of the best-fit image ellipse. Measures the longitudinal development of the shower image. \\
DENS & A measure of image compactness, defined as $log_{10}(SIZE / (WIDTH \times LENGTH))$. Gamma-ray images are typically denser. \\
CONC & Image concentration, the ratio of the charge in the two brightest pixels to the total \texttt{SIZE}. \\
LEAKAGE & The fraction of the total charge contained in pixels at the edge of the camera. Used to reject events that are not fully contained in the field of view. \\
NUSEDTEL & The number of telescopes contributing to the stereoscopic reconstruction. A higher number generally leads to better reconstruction quality. \\
TELIP & The reconstructed impact parameter; the distance between the shower core on the ground and a given telescope's position. \\
STMAXH & The reconstructed height of the shower maximum above sea level. Hadronic showers tend to develop higher in the atmosphere. \\
\noalign{\smallskip}\hline\noalign{\smallskip}
\end{tabular}
\end{table}

\subsection{The Random Forest: A Powerful and Interpretable Baseline}
\label{subsec:3.3}

For many years, the Random Forest (RF) algorithm has been the workhorse of IACT data analysis and serves as a powerful and robust baseline model~\cite{Albert:2008gq, Ohm:2009, Pagliaro:2023}. To understand the RF, it is helpful to first introduce its building block: the \emph{decision tree}. A decision tree is a simple predictive model that works by asking a sequence of binary questions about the input features---for example, ``Is the image \texttt{Width} less than 0.05 degrees?''---and following the answers down a tree-like structure of branches until a final prediction (a class label or a numerical value) is reached at a leaf node~\cite{Bishop:2006}. Each individual decision tree is easy to interpret but tends to be unstable: small changes in the training data can produce very different trees, and a single tree can easily \emph{overfit} the data---that is, learn the noise and statistical fluctuations in the training set rather than the genuine underlying patterns, leading to poor performance on new, unseen data~\cite{Hastie:2009}.

An RF overcomes these limitations by constructing a large number of decision trees and combining their outputs: it takes the majority vote for classification or the average prediction for regression~\cite{Breiman:2001}. Its effectiveness stems from two key principles:

\begin{enumerate}
    \item \textbf{Bootstrap Aggregating (Bagging):} Each individual decision tree in the forest is trained on a different random subsample of the training data, drawn with replacement. This process ensures that the individual trees are diverse and reduces the overall model's variance, making it less prone to overfitting~\cite{Breiman:2001}.
    \item \textbf{Random Subspace Method:} When splitting a node in a tree, the algorithm considers only a random subset of the available features. This decorrelates the trees from one another, as it prevents a single, highly predictive feature from dominating the structure of all trees in the forest~\cite{Breiman:2001}.
\end{enumerate}

\begin{figure}
\centering 
\includegraphics[width=5 cm]{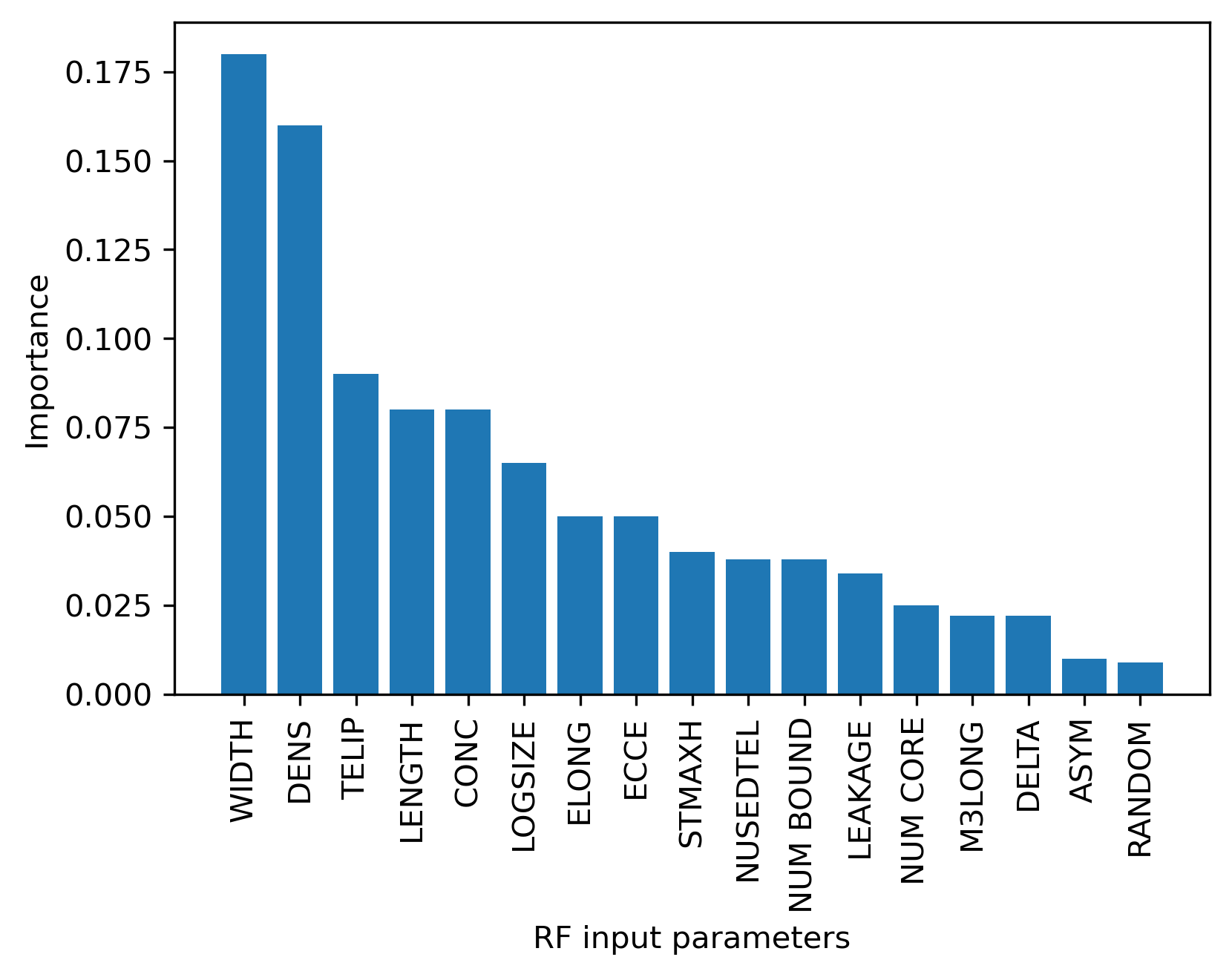}
\includegraphics[width=5 cm]{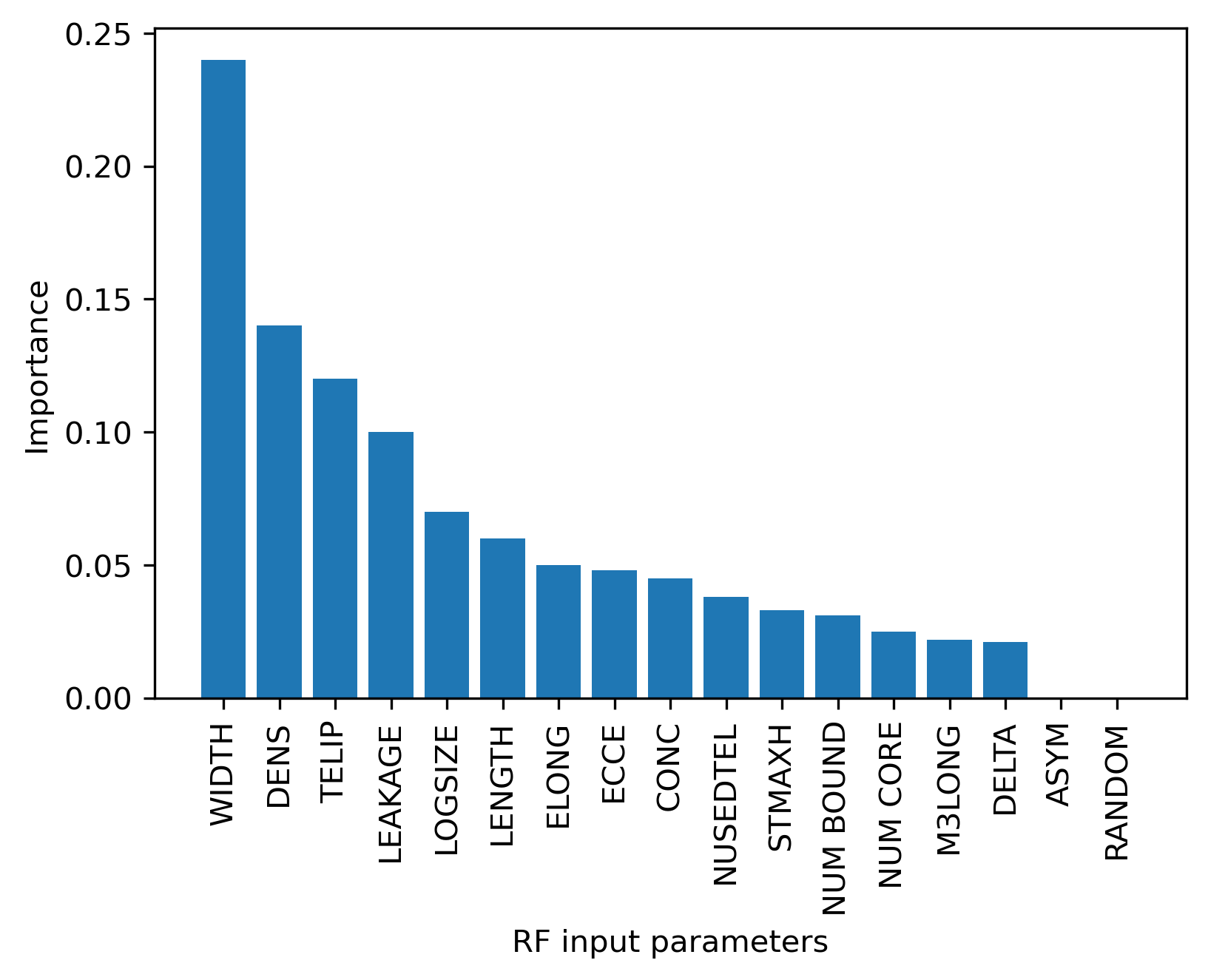}
\caption{Feature importances for gamma/hadron separation computed with two methods:\linebreak (\textbf{Left}) Mean decrease in impurity method. (\textbf{Right}) Permutation~method. Reprinted from \cite{Pagliaro:2023} licensed under CC BY 4.0. \label{fig:gh_feat_importance}}
\end{figure}

\begin{figure}
\centering 
\includegraphics[width=5 cm]{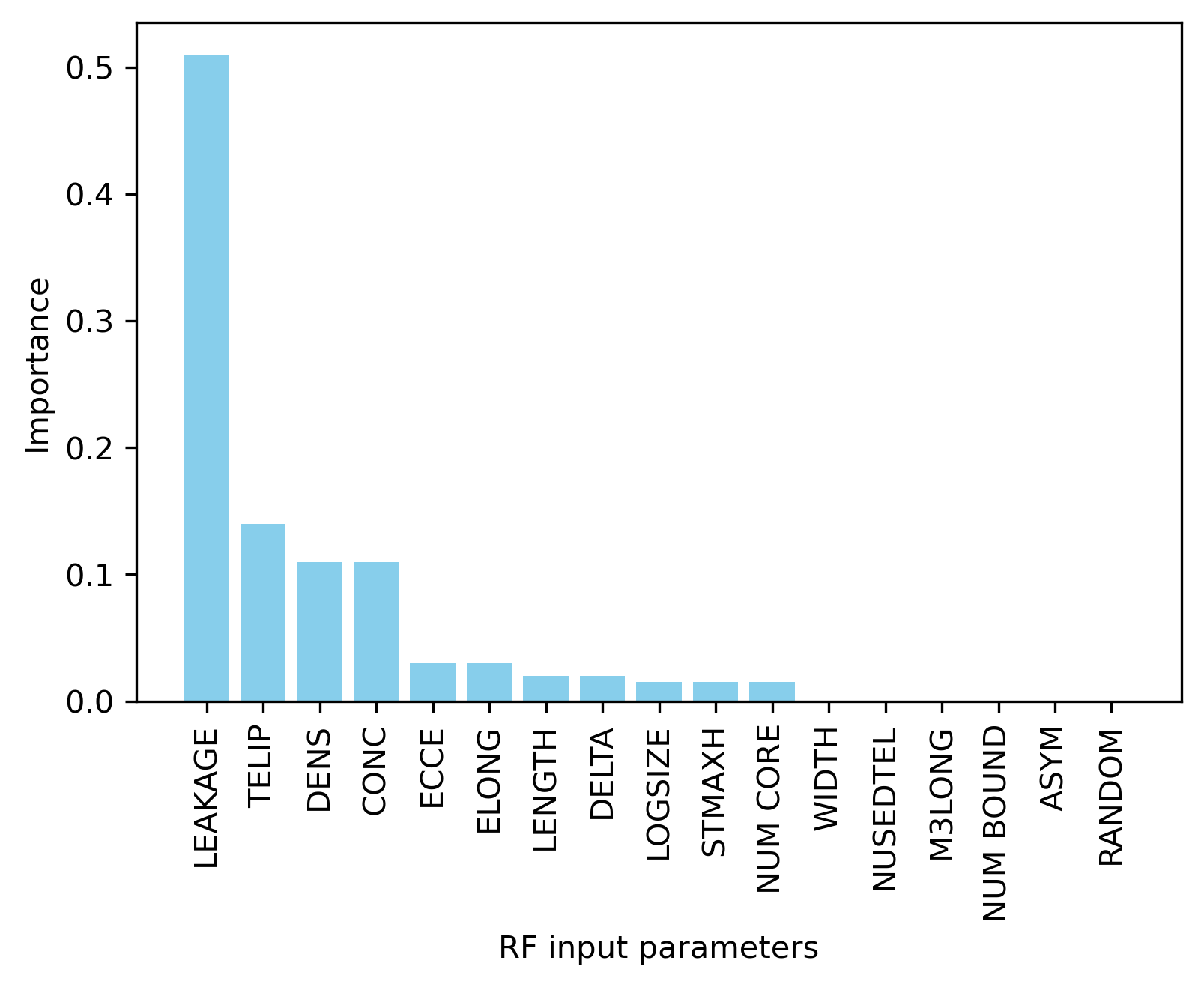}
\includegraphics[width=5 cm]{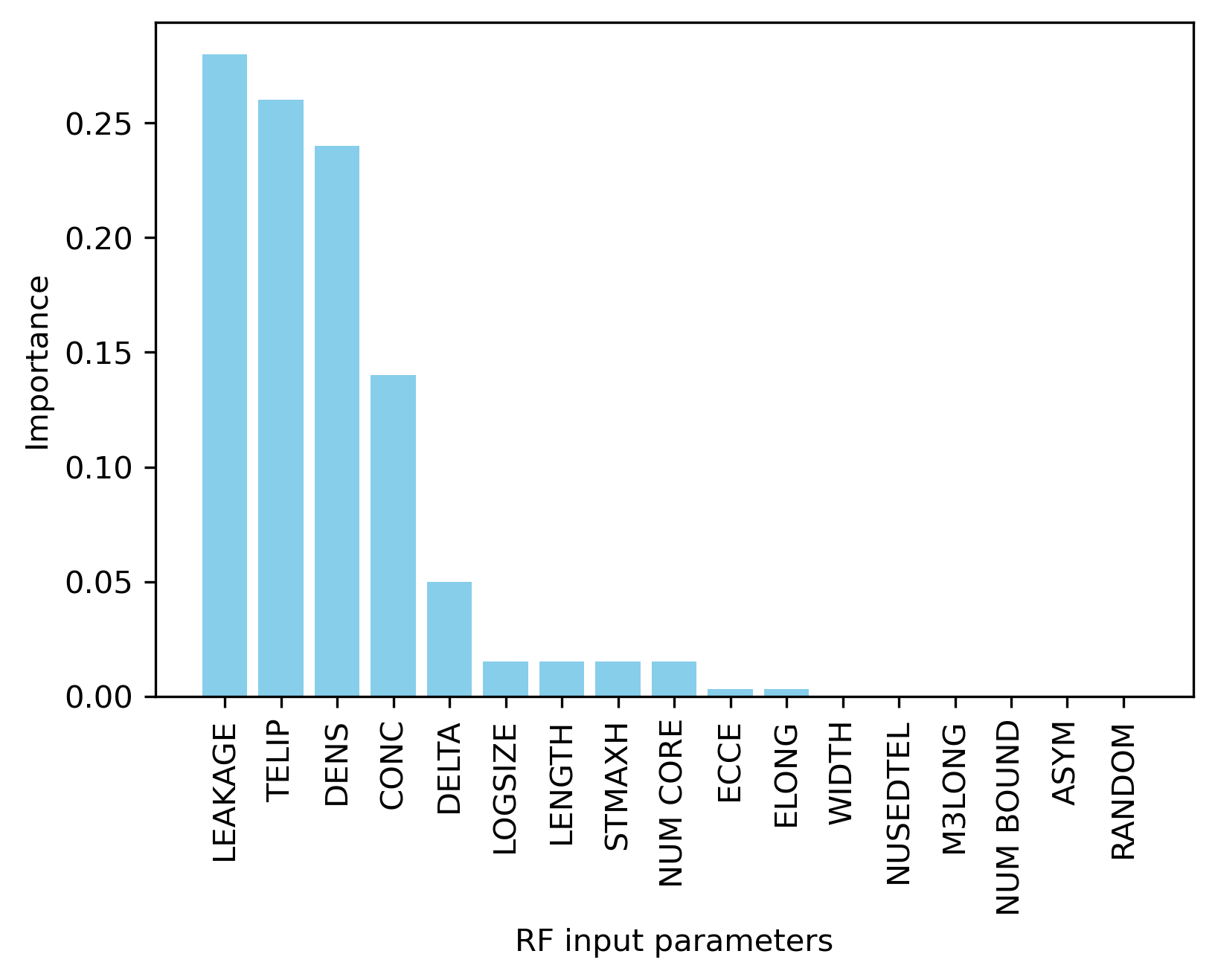}
\caption{Feature importances for energy reconstruction computed with two methods: (\textbf{Left}) Mean decrease in impurity method. (\textbf{Right}) Permutation~method. Reprinted from \cite{Pagliaro:2023} licensed under CC BY 4.0. \label{fig:energy_feat_importance}}
\end{figure}

The combination of many diverse, decorrelated ``weak learners''---models that individually perform only slightly better than random guessing---creates a single ``strong learner'' (the forest) whose collective accuracy far exceeds that of any individual tree. This resulting ensemble is robust, handles high-dimensional and complex feature spaces well, and is relatively insensitive to hyperparameter tuning---that is, the choices made by the analyst before training, such as the number of trees in the forest or the maximum depth of each tree, have a limited impact on the final performance compared to other algorithms~\cite{Breiman:2001}. A particularly valuable feature of RFs is their ability to provide a measure of feature importance. Two complementary metrics are shown in Figs.~\ref{fig:gh_feat_importance}--\ref{fig:energy_feat_importance}: the Mean Decrease in Impurity (MDI), which accumulates the reduction in node impurity across all splits, and the permutation importance, which measures the drop in predictive performance when a feature's values are randomly shuffled. The relative merits and limitations of these metrics---and a more rigorous alternative---are discussed in Section~\ref{subsec:3.4}.

\subsection{Feature Importance: From Native Metrics to Ablation Studies}
\label{subsec:3.4}

The Random Forest provides, as a by-product of training, a ranking of how
much each input feature contributes to its decisions. Understanding this
ranking is valuable both for physical interpretation---confirming that the
model exploits physically meaningful patterns---and for practical feature
selection. However, the meaning of ``feature importance'' is not unique:
different metrics answer subtly different questions, and conflating them
can lead to incorrect conclusions. This section introduces the three most
widely used approaches (Gini importance, permutation importance, and SHAP
values), explains their limitations, and advocates for \emph{ablation
studies} as a more rigorous alternative.

\subsubsection{Gini Importance (Mean Decrease Impurity)}
\label{subsubsec:mdi}

The most readily available metric is the \emph{Mean Decrease in Impurity}
(MDI), also called Gini importance~\cite{Louppe:2013}. At each node of each
decision tree, the algorithm selects the feature and threshold that most
reduces the impurity (typically measured by the Gini index or Shannon
entropy) of the resulting child nodes. Summing these reductions across all
nodes that split on feature $f_j$, weighted by the fraction of training
samples $n_t / N$ reaching each node $t$, and averaging over all trees,
gives:
\begin{equation}
  I^{\text{MDI}}_j \;=\; \frac{1}{T}\sum_{k=1}^{T}
  \sum_{\substack{t \,\in\, \text{tree}_k \\ \text{split on }f_j}}
  \frac{n_t}{N}\,\Delta \text{impurity}(t)\,,
  \label{eq:mdi}
\end{equation}
where $T$ is the number of trees and $N$ is the total number of training
samples. MDI is computed for free during training and is the metric shown
in the left panels of Figs.~\ref{fig:gh_feat_importance} and~\ref{fig:energy_feat_importance}. Its main
weakness is that it measures model-\emph{internal} behaviour rather than
actual predictive performance on held-out data. It is also systematically
biased towards features with a large number of possible split points
(e.g., continuous features with high dynamic range), and it can overstate
the importance of redundant features when several correlated inputs are
available~\cite{Strobl:2007}.

\subsubsection{Permutation Importance (Mean Decrease Accuracy)}
\label{subsubsec:mda}

A more reliable alternative evaluates feature relevance on the held-out
test set. The \emph{permutation importance} of feature $f_j$ is measured
by randomly shuffling its values across test events, breaking the
statistical relationship between $f_j$ and the target variable, and
recording the resulting drop in performance~\cite{Breiman:2001}:
\begin{equation}
  \text{PI}_j \;=\; \text{score}_{\text{baseline}}
               \;-\; \text{score}_{\text{shuffled }j}\,.
  \label{eq:pi}
\end{equation}
Unlike MDI, permutation importance is tied to actual predictive performance
and can be applied to any model. Its results are shown in the right panels
of Figs.~\ref{fig:gh_feat_importance} and~\ref{fig:energy_feat_importance}. However, it has a subtle flaw
when features are correlated~\cite{Hooker:2021}: shuffling $f_j$ independently
of the other features generates event records where the \emph{joint}
distribution of features is unrealistic---combinations that never appear in
nature. The model is effectively evaluated on out-of-distribution inputs,
which can produce misleading scores. In IACT analyses, where parameters
such as $\texttt{DENS} = \log_{10}(\texttt{SIZE}/(\texttt{WIDTH} \times
\texttt{LENGTH}))$ are algebraically related to others, this extrapolation
effect is non-negligible.

\subsubsection{SHAP Values}
\label{subsubsec:shap}

SHAP (SHapley Additive exPlanations) is grounded in cooperative game
theory~\cite{Lundberg:2017}. The SHAP value of feature $f_j$ for a single
prediction is the average marginal contribution of $f_j$ across all
possible orderings in which features are introduced:
\begin{equation}
  \phi_j \;=\;\sum_{S \,\subseteq\, F \setminus \{j\}}
  \frac{|S|!\,(|F|-|S|-1)!}{|F|!}\,
  \bigl[f(S \cup \{j\}) - f(S)\bigr]\,,
  \label{eq:shap}
\end{equation}
where $F$ is the full feature set and $f(S)$ is the model output when only
the features in subset $S$ are available. This formulation satisfies
axiomatic fairness properties: the individual SHAP values sum exactly to
the difference between the model's prediction and the global mean, and the
assignment is unique. For tree-based models, the \texttt{TreeSHAP}
algorithm computes exact Shapley values in polynomial time, making it
practical for the datasets typical of IACT analyses~\cite{Lundberg:2017}.
Global importance is then summarised as the mean $|\phi_j|$ over all test
events. The key caveat is shared with permutation importance: SHAP explains
\emph{what the trained model does}, not \emph{how much performance degrades
if a feature is absent}.

\subsubsection{Why Ablation Studies Are Necessary}
\label{subsubsec:ablation_motivation}

All three metrics above are \emph{post-hoc} analyses of an already-trained
model: they answer the question ``which features does this model rely on?''
An ablation study asks a different and more practically relevant question:
``how much would performance degrade if this feature were unavailable?''

The protocol is straightforward. Given a baseline model trained on the
full feature set $\mathcal{F}$ with performance $Q_{\text{baseline}}$,
each feature $f_j$ is ablated by \emph{retraining} the model on
$\mathcal{F} \setminus \{f_j\}$ and measuring:
\begin{equation}
  \Delta Q_j \;=\; Q_{\text{baseline}} - Q_{\setminus f_j}\,.
  \label{eq:ablation}
\end{equation}
A positive $\Delta Q_j$ confirms that $f_j$ genuinely improves performance;
a negative value indicates it adds noise. Crucially, because the model is
\emph{retrained} without $f_j$, it is never evaluated on out-of-distribution
inputs: the extrapolation problem of permutation importance is avoided by
design~\cite{Hooker:2021}. The results are also expressed in units of the
science metric $Q$, making it immediately clear whether a feature's
contribution is physically meaningful or merely a statistical artefact.
The computational cost---one full train-test cycle per feature---is
entirely tractable for the $\mathcal{O}(10)$ feature sets typical of IACT
analyses. Table~\ref{tab:importance_comparison} summarises the key
properties of all four approaches.

\begin{table}[t]
\caption{Comparison of four approaches to feature importance. The first three
analyse an already-trained model; ablation requires retraining but provides
a direct, out-of-distribution-free measure of end-to-end performance impact.}
\label{tab:importance_comparison}
\begin{tabular}{p{2.1cm}p{3.2cm}p{1.6cm}p{3.4cm}}
\hline\noalign{\smallskip}
Method & What it measures & Retraining needed & Key limitation \\
\noalign{\smallskip}\svhline\noalign{\smallskip}
Gini (MDI)        & Node impurity reduction during training & No
  & Biased to high-cardinality features; ignores correlations \\[4pt]
Permutation       & Performance drop on shuffled test set & No
  & Creates out-of-distribution inputs when features are correlated \\[4pt]
SHAP              & Marginal Shapley contribution per prediction & No
  & Explains model predictions, not performance impact \\[4pt]
Ablation          & Performance drop after retraining without feature & Yes
  & Computationally more expensive \\
\noalign{\smallskip}\hline\noalign{\smallskip}
\end{tabular}
\end{table}

\subsubsection{Case Study: MC Ablation for the ASTRI Mini-Array}
\label{subsubsec:ablation_astri}

To illustrate the ablation methodology in a concrete setting, we present a
systematic feature ablation study on Monte Carlo data for the ASTRI
Mini-Array. The results shown here are preliminary and based on an analysis
in preparation~\cite{Pagliaro:2025}. Three classifiers are evaluated---Random Forest (RF),
ExtraTrees (ET), and LightGBM---in both mono (single-telescope, 11
features) and stereo (multi-telescope, 15 features) reconstruction modes,
using the same MC dataset described in Section~\ref{subsec:3.2} and an
80/20 stratified train/test split. The full feature sets are those from
Tables~\ref{tab:morph_params} and~\ref{tab:temp_params}. The baseline
Quality Factor $Q_{\text{baseline}}$%
\footnote{The Quality Factor $Q = \epsilon_\gamma / \sqrt{\epsilon_{\text{bkg}}}$
is the standard figure of merit for gamma/hadron separation and acts as a proxy
for the statistical significance of a detection; see Section~\ref{subsec:6.1}
for a full discussion.}
is established on the held-out test
set, and then each feature is ablated following Eq.~(\ref{eq:ablation}).
Results are reported in Table~\ref{tab:ablation_mc}.

\textbf{\textsc{Width} is the single most discriminating feature.} Across
all configurations, removing \textsc{width} produces the largest performance
drop. In mono mode with RF, $\Delta Q = +1.03$, a 25\% degradation from
$Q_0 = 4.04$. This is physically well-motivated: the transverse width of
the Cherenkov image is the most direct geometric tracer of the lateral
extent of the particle cascade, which is far narrower for electromagnetic
(gamma-ray) showers than for hadronic ones.

\textbf{The ranking is model-dependent.} A revealing result in stereo mode
is that the feature hierarchy depends significantly on the classifier
architecture. For RF, \textsc{width} remains dominant ($\Delta Q = +1.44$).
For LightGBM, however, \textsc{dist}---the reconstructed angular impact
distance---becomes the most critical feature ($\Delta Q = +1.02$), while
\textsc{width} drops to $+0.13$. This occurs because LightGBM's sequential
gradient boosting strategy and RF's parallel bagging learn complementary
decompositions of the feature space. Neither is more ``correct''; they
exploit the available information differently. A closely related
manifestation is visible in the ExtraTrees baseline, whose mono
$Q_0 = 2.53$ lags well behind RF (4.04) and LightGBM (4.28): extreme
split randomization is an asset when discriminative information is
spread across many correlated features---as in the 15-feature stereo
setup, where the gap narrows---but a liability in mono, where
discriminative power is concentrated in a few dominant variables
(\textsc{width}, \textsc{leakage}) that are better exploited by
deterministic splitting. These observations confirm the argument developed above:
native importance scores are intrinsic to a specific architecture and
cannot be compared across classifiers without the regularizing frame of
an ablation study.

\textbf{\textsc{Conc} is negligible or slightly harmful on MC.} The
concentration parameter---fraction of charge in the two brightest
pixels---shows $\Delta Q \leq 0$ across all MC configurations. This
indicates that on idealised simulation data \textsc{conc} carries no
information not already captured by the shape parameters, and combined
with its high anti-correlation with \textsc{length} ($r = -0.79$), it
acts as a noise source. This result motivates real-data validation:
whether this conclusion holds when instrumental effects such as
point-spread-function broadening, mirror ageing, and night-sky background
modify the pixel-level light distribution is an empirical question that
ablation on observed data can directly address.

\textbf{\textsc{Dens} is partially redundant.} The derived parameter
$\texttt{DENS} = \log_{10}(\texttt{SIZE}/(\texttt{WIDTH} \times
\texttt{LENGTH}))$ shows modest contributions ($\Delta Q \lesssim 0.3$)
that are fully replaceable by its constituent features individually---a
consequence of the high algebraic correlation with $\log_{10}(\textsc{size})$
($r = 0.85$). Permutation importance would overstate its relevance by
treating it as independent; ablation exposes the redundancy directly.

\begin{table}[t]
\caption{Single-feature ablation study on Monte Carlo ASTRI Mini-Array
simulations. $\Delta Q_j = Q_{\text{baseline}} - Q_{\setminus f_j}$:
positive values indicate beneficial features; negative values indicate
noise. Results shown for RF, ExtraTrees (ET), and LightGBM in mono
(baseline $Q_0$: RF\,=\,4.04, ET\,=\,2.53, LGB\,=\,4.28) and stereo
($Q_0$: RF\,=\,5.79, ET\,=\,3.38, LGB\,=\,6.73) modes. Only features
with $|\Delta Q| > 0.05$ for at least one classifier are shown.}
\label{tab:ablation_mc}
\small
\begin{tabular}{lrrr|lrrr}
\hline\noalign{\smallskip}
\multicolumn{4}{c|}{Mono (11 features)} & \multicolumn{4}{c}{Stereo (15 features)} \\[2pt]
Feature & \small$\Delta Q_\text{RF}$ & \small$\Delta Q_\text{ET}$ & \small$\Delta Q_\text{LGB}$
        & Feature & \small$\Delta Q_\text{RF}$ & \small$\Delta Q_\text{ET}$ & \small$\Delta Q_\text{LGB}$ \\
\noalign{\smallskip}\svhline\noalign{\smallskip}
\textsc{width}          & $+1.03$ & $+0.44$ & $+0.41$  &
  \textsc{width}        & $+1.44$ & $+0.48$ & $+0.13$  \\
\textsc{leakage}        & $+0.32$ & $+0.00$ & $+0.39$  &
  \textsc{dist}         & $+0.58$ & $+0.25$ & $+1.02$  \\
$\log_{10}$\textsc{size}& $+0.12$ & $+0.14$ & $-0.00$  &
  \textsc{leakage}      & $+0.45$ & $+0.23$ & $+0.59$  \\
\textsc{dens}           & $+0.07$ & $+0.26$ & $+0.02$  &
  \textsc{telip}        & $+0.19$ & $+0.04$ & $+0.30$  \\
\textsc{length}         & $+0.07$ & $+0.05$ & $+0.19$  &
  \textsc{dens}         & $+0.19$ & $+0.10$ & $+0.13$  \\
\textsc{conc}           & $-0.04$ & $-0.15$ & $+0.02$  &
  \textsc{nusedtel}     & $+0.11$ & $+0.20$ & $+0.26$  \\
\textsc{asym}           & $-0.06$ & $-0.16$ & $+0.02$  &
  \textsc{stmaxh}       & $+0.02$ & $+0.25$ & $+0.11$  \\
\noalign{\smallskip}\hline\noalign{\smallskip}
\end{tabular}
\end{table}

\subsection{Analysis Software Frameworks}
\label{subsec:3.5}

The machine learning techniques described above are not implemented in isolation but are embedded within comprehensive software frameworks that manage the full analysis chain, from raw data calibration to high-level scientific products. Each major IACT experiment has developed its own framework. The MAGIC collaboration uses the \textbf{MARS} (MAGIC Analysis and Reconstruction Software) pipeline~\cite{Albert:2008gq}, which includes a Random Forest implementation for both gamma/hadron separation and energy estimation. H.E.S.S.\ relies on several parallel analysis chains, including one based on a semi-analytical shower model and another employing multivariate methods~\cite{Ohm:2009, deNaurois:2015}. VERITAS uses the \textbf{EventDisplay} and \textbf{VEGAS} packages for its standard analyses~\cite{Holder:2012}.

Looking toward the next generation, the CTAO project has driven the development of open-source, community-wide tools. \textbf{ctapipe}~\cite{ctapipe:2024} is a Python library that provides the low-level building blocks for IACT data processing: calibration, image cleaning, Hillas parameterization, and stereoscopic reconstruction. It is designed to be instrument-agnostic and extensible, serving as the foundation for the official CTAO analysis pipeline. Built on top of ctapipe, \textbf{lstchain} provides the analysis pipeline specific to the Large-Sized Telescopes, including machine-learning-based reconstruction. For high-level science analysis---spectral fitting, light-curve extraction, sky-map generation---\textbf{Gammapy}~\cite{Gammapy:2023} has emerged as the reference tool. Gammapy implements the standard statistical methods of the field (e.g., the Li \& Ma significance test, forward-folding spectral analysis) in a unified, open-source framework that can ingest data from any IACT experiment, promoting reproducibility and cross-experiment comparisons.

The convergence toward open, standardized software is an important development for the field, as it enables the broader astrophysics community---including researchers from adjacent disciplines---to access and analyze IACT data.

\section{Pushing the Frontiers I: The Temporal Dimension}
\label{sec:4}

While the morphological and stereoscopic parameters provide a powerful basis for reconstruction, they discard a rich source of information: the arrival time of the Cherenkov photons at the camera. Modern IACT cameras can record not just the integrated charge in each pixel, but also a precise time tag for when the signal arrived. This temporal dimension opens a new frontier for feature engineering, allowing for significant enhancements in performance.

\subsection{The Physical Motivation for Using Timing Information}
\label{subsec:4.1}

The time evolution of a shower image contains profound physical information. A gamma-ray shower develops longitudinally along a well-defined axis in the atmosphere. Cherenkov photons emitted from higher altitudes must travel further to reach the telescope, arriving slightly later than photons emitted from lower altitudes. This creates a smooth and predictable time gradient across the image on the camera plane, with one end of the ellipse lighting up before the other~\cite{LaParola:2025}.

Hadronic showers, in contrast, have a more complex temporal structure. They are spatially broader, and the presence of secondary muons, which travel at nearly the speed of light with little interaction, can produce distinct, early-arriving signals far from the shower core. The overall temporal development is less coherent and more chaotic than that of a gamma-ray shower~\cite{LaParola:2025}. This fundamental physical difference can be exploited for gamma/hadron separation.

The use of timing information is not merely a theoretical concept; it has been successfully implemented by major IACT experiments. The MAGIC telescopes, for example, incorporated temporal constraints into their image cleaning and used timing parameters like time gradient and root mean square (RMS) to halve the residual hadronic background in their analysis~\cite{Aliu:2009, Albert:2008gq}. Similar efforts by the VERITAS and H.E.S.S. collaborations have also demonstrated the efficacy of this approach~\cite{LaParola:2025}.

\begin{figure}
\centering
    \includegraphics[width=10cm]{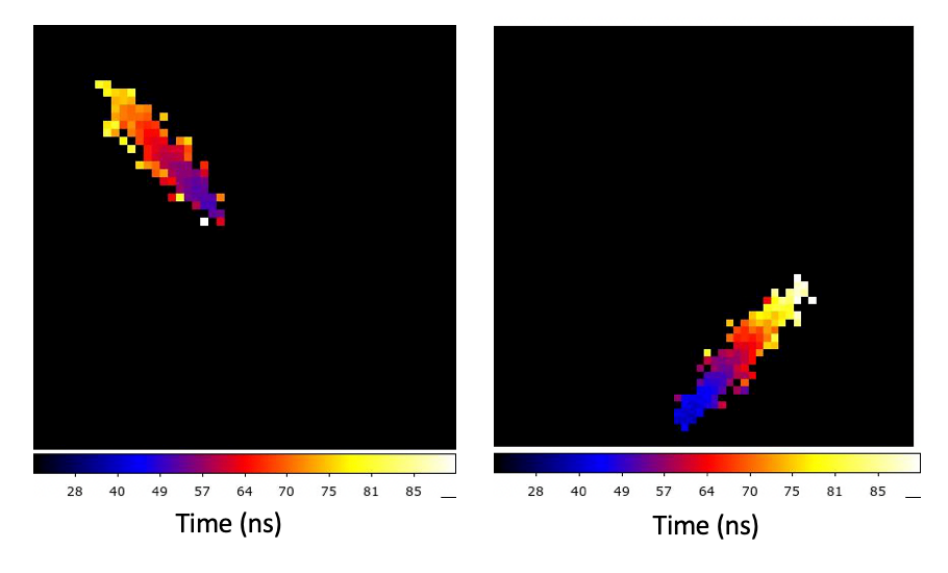} \\
\caption{Simulated Cherenkov shower images as observed by an ASTRI Mini-Array telescope, illustrating the differences in time evolution between different types of events. The images cover the entire ASTRI field of view (each side spanning approximately 10 deg). The color scale indicates the time when the pixel is triggered by the incoming Cherenkov photons, {\bf with the black indicating that the pixel has not been triggered}. The time (in ns) increases from blue to white. In this example, the gamma-initiated event (\textbf{Left}) evolves more slowly than the hadron-initiated event (\textbf{Right}). Reprinted from \cite{LaParola:2025} licensed under CC BY 4.0. }
\label{fig:astri_time_evolution}
\end{figure}   

\begin{figure}
    \centering
    \includegraphics[width=10cm]{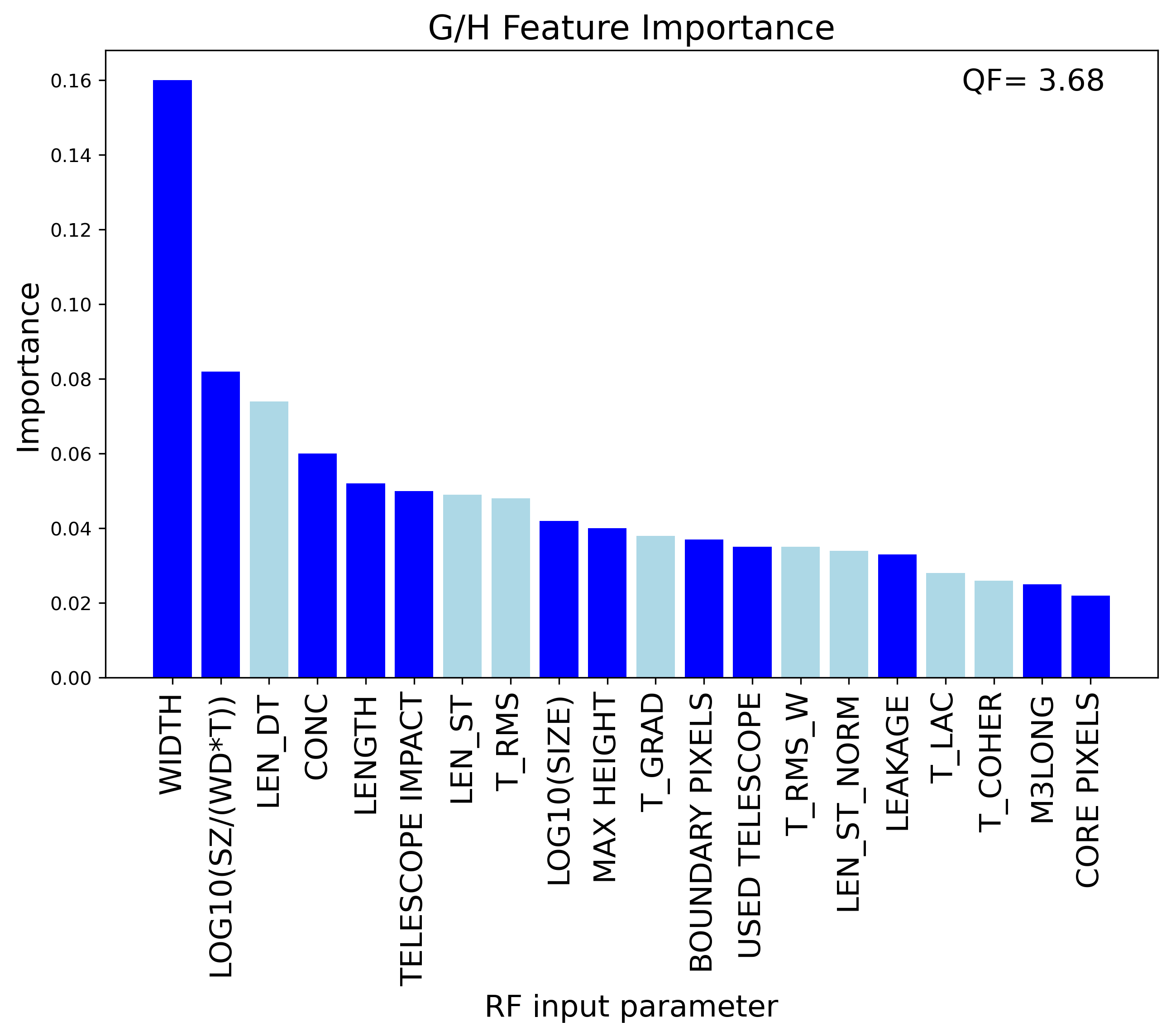}
    \caption{Feature importance of morphological, stereo, and~selected temporal parameters using random~forest (blue for morphological and stereo parameters, light blue for temporal parameters). Reprinted from \cite{LaParola:2025} licensed under CC BY 4.0. }
    \label{fig:feature_importance_selected}
\end{figure}

\subsection{A Lexicon of Temporal Parameters: From Dispersion to Coherence}
\label{subsec:4.2}

Building on this physical motivation, a comprehensive set of temporal parameters can be engineered to capture different aspects of a shower's time evolution. The following parameters, developed in the context of the ASTRI Mini-Array, provide a powerful lexicon for describing the temporal structure of a Cherenkov image~\cite{LaParola:2025}. Let $t_p$ be the time recorded in a given pixel, $i_p$ be its charge, and $n_p$ be the total number of pixels in the cleaned image.

\begin{table}
\caption{Key Temporal Parameters for IACT Event Reconstruction. These parameters capture the timing structure of the Cherenkov shower, providing information complementary to the standard morphological features.}
\label{tab:temp_params}       
\begin{tabular}{p{2.5cm}p{9cm}}
\hline\noalign{\smallskip}
Parameter & Physical Interpretation  \\
\noalign{\smallskip}\svhline\noalign{\smallskip}
$T_{RMS}$ & The Root Mean Square of the pixel arrival times. Measures the overall duration or temporal spread of the shower signal. \\
$T_{GRAD}$ & The linear gradient of arrival times along the image's major axis. Measures how quickly the shower front sweeps across the camera. \\
$T_{RMS\_W}$ & A charge-weighted version of $T_{RMS}$. Gives more importance to the timing of brighter parts of the image. \\
$LEN_{ST}$ & Length in "Size-Time" space. Captures the correlation between pixel brightness and arrival time. \\
$LEN_{DT}$ & Length in "Distance-Time" space. Captures the correlation between a pixel's distance from the image center and its arrival time. \\
$T_{LAC}$ & Lacunarity of the arrival time sequence. Quantifies the "gappiness" or temporal inhomogeneity of the photon arrivals. Hadronic showers are expected to be more lacunar. \\
$T_{COHER}$ & Temporal Coherence. Measures the monotonicity of the shower's spatial development over time. Gamma-ray showers are expected to be more coherent. \\
\noalign{\smallskip}\hline\noalign{\smallskip}
\end{tabular}
\end{table}

The mathematical definitions for some of these key parameters are as follows. The standard time RMS is defined relative to the average pixel time, $t_{AVG} = \frac{\Sigma t_p}{n_p}$:
\begin{equation}
T_{RMS} = \sqrt{\sum(t_p - t_{AVG})^2}
\label{eq:trms}
\end{equation}
The weighted time RMS uses a charge-weighted average time, $t_{AVGW} = \frac{\Sigma(t_p \times i_p)}{\Sigma i_p}$, and is defined as:
\begin{equation}
T_{RMS\_W} = \sqrt[4]{\frac{\sum(t_p - t_{AVGW})^2}{n_p^3}}
\label{eq:trmsw}
\end{equation}
We follow here the notation and conventions of~\cite{LaParola:2025}. Despite the ``RMS'' label, these expressions are not normalized as standard root-mean-square dispersions: Eq.~(\ref{eq:trms}) omits the $1/n_p$ normalization and Eq.~(\ref{eq:trmsw}) uses a fourth root rather than a square root---a choice motivated in the original analysis by the enhanced gamma/hadron separation power it yields. Both should therefore be understood as engineered discriminant variables rather than statistical moments.

Lacunarity is a concept originally introduced in fractal geometry to quantify the ``gappiness'' or inhomogeneity of a spatial or temporal pattern~\cite{LaParola:2025}. A perfectly uniform signal has a lacunarity of 1, while increasingly irregular or clustered signals have higher values. In this context, it is applied to the sequence of photon arrival times: a gamma-ray shower, with its smooth and coherent temporal development, is expected to produce a more uniform (lower lacunarity) time sequence than a hadronic shower, whose temporal structure is fragmented by the multiple sub-showers. Lacunarity is calculated using the gliding box method on a discretized timeline. It is defined by the mean and variance of the probability distribution $P_M$ of the number of triggered time bins (mass, M) within the gliding box:
\begin{equation}
T_{LAC} = \frac{\mathrm{Var}(P_M)}{\mathrm{Avg}(P_M)^2} + 1
\label{eq:tlac}
\end{equation}
These parameters, summarized in Table \ref{tab:temp_params}, provide a rich, multi-faceted description of the shower's temporal dynamics.

\subsection{Case Study: Enhancing Gamma/Hadron Separation with Temporal Features}
\label{subsec:4.3}

The practical value of these temporal parameters has been demonstrated through detailed analysis of simulated data for the ASTRI Mini-Array~\cite{LaParola:2025}. By adding a carefully selected set of eight temporal parameters (including $T_{RMS}$, $T_{GRAD}$, $T_{LAC}$, and $T_{COHER}$) to the standard set of morphological and stereoscopic features, a significant improvement in gamma/hadron separation was achieved.

The performance was quantified using the Quality Factor (QF), a standard metric in the field (defined in Section \ref{sec:6}). The inclusion of temporal features increased the QF of the Random Forest classifier from 3.32 to 3.68, representing a performance gain of approximately 10.8\%~\cite{LaParola:2025}.

Crucially, this improvement is not uniform across all energies. The analysis revealed that the temporal parameters have their most profound impact at lower energies, below approximately 1.5 TeV. In this regime, the inclusion of timing information can improve the array's sensitivity by up to 50\%~\cite{LaParola:2025}. This finding is not merely a technical detail; it has deep scientific implications. The scientific frontiers for next-generation IACTs---transient events like gamma-ray bursts, distant active galactic nuclei, and any source whose spectrum cuts off below the TeV---depend critically on sensitivity at the lowest accessible energies, where showers are fainter, less developed, and morphologically ambiguous. Traditional Hillas parameters lose discriminating power in this regime~\cite{Shilon:2018}, and it is precisely here that temporal features contribute the most. This is a concrete example of how scientific requirements shape the direction of methodological work in the field, rather than the other way around.

\section{Pushing the Frontiers II: Advanced Ensemble Learning}
\label{sec:5}

In parallel with the development of more sophisticated features, an active area of research explores the use of more advanced machine learning models. It should be emphasized that, at the time of writing, the Random Forest remains the algorithm of choice in the standard analysis pipelines of all major operating IACT experiments. The techniques discussed in this section are the subject of ongoing studies aimed at assessing whether they can offer superior performance, particularly for the regression task of energy reconstruction.

\subsection{Beyond Random Forests: Gradient Boosting and Extra Trees}
\label{subsec:5.1}

Two powerful alternatives to the Random Forest algorithm are Extreme Gradient Boosting (XGBoost) and Extra Trees.

\begin{itemize}
    \item \textbf{Extreme Gradient Boosting (XGBoost):} Unlike the parallel "bagging" approach of Random Forest, XGBoost~\cite{Chen:2016} uses a sequential ``boosting'' technique. It builds an ensemble of decision trees one at a time, where each new tree is trained specifically to correct the errors made by the previous trees in the ensemble. This iterative process works by minimizing a \emph{loss function}---a mathematical quantity that measures how far the model's predictions are from the true values---using a technique called \emph{gradient descent}, in which the algorithm adjusts the model parameters step by step in the direction that most rapidly reduces the loss~\cite{Bishop:2006}. The result is often a model with very high predictive accuracy~\cite{Pagliaro:2023}.
    \item \textbf{Extra Trees (Extremely Randomized Trees):} This method is similar to Random Forest but introduces an additional layer of randomization. When choosing a split point for a feature, instead of searching for the optimal threshold, it selects a random threshold. This can further reduce the model's variance and computational cost, sometimes leading to better generalization~\cite{Pagliaro:2023}.
\end{itemize}

\subsection{Stacking Ensembles: Combining Models for Optimal Performance}
\label{subsec:5.2}

Different machine learning algorithms have different strengths and weaknesses, and may make different types of errors. An intuitive analogy is that of a panel of experts: each expert (model) has a different area of competence, and a moderator (the meta-learner) listens to all of their opinions and learns, from experience, how much weight to give to each expert depending on the circumstances. Stacking formalizes this intuition as a meta-learning technique that combines the strengths of multiple base models to create a more powerful final model.

To illustrate the potential of this approach, we summarize here the stacking architecture explored in~\cite{Pagliaro:2023} as a concrete example of what can be achieved. The process involves two levels of learning. First, several diverse base models (e.g., a Random Forest, an XGBoost model, and an Extra Trees model) are trained on the full training dataset. Then, a second-level model, or "meta-learner," is trained. The input features for this meta-learner are the predictions made by the base models. The meta-learner's task is to learn the optimal way to combine the predictions from the base models to produce the final, improved prediction~\cite{Pagliaro:2023}. For instance, it might learn that the XGBoost model is most trustworthy for high-energy events, while the Random Forest performs better for low-energy events.

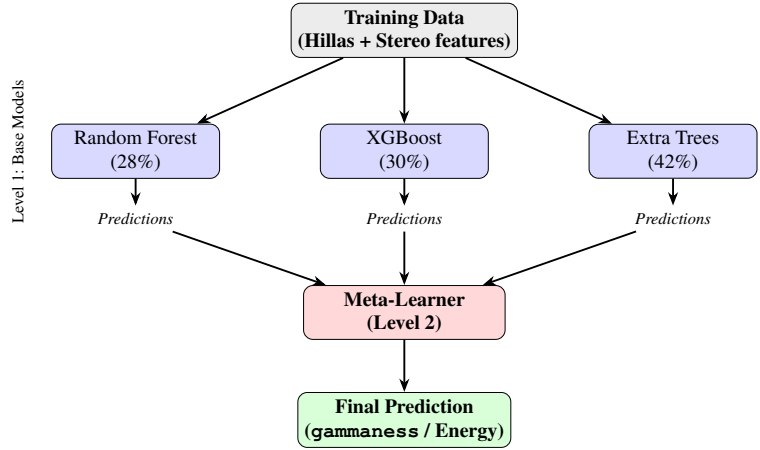
\begin{figure}[t]
\centering
\resizebox{0.85\textwidth}{!}{%
\begin{tikzpicture}[
    every node/.style={font=\small},
    data/.style={draw, rounded corners, fill=gray!15, minimum width=3cm, minimum height=0.7cm, align=center, font=\small\bfseries},
    model/.style={draw, rounded corners, fill=blue!15, minimum width=2.5cm, minimum height=0.7cm, align=center},
    meta/.style={draw, rounded corners, fill=red!15, minimum width=3cm, minimum height=0.7cm, align=center, font=\small\bfseries},
    output/.style={draw, rounded corners, fill=green!15, minimum width=3cm, minimum height=0.7cm, align=center, font=\small\bfseries},
    arr/.style={-{Stealth[length=2mm]}, thick}
]
\node[data] at (4,0) (train) {Training Data\\(Hillas + Stereo features)};

\node[model] at (0,-1.8) (rf) {Random Forest\\(28\%)};
\node[model] at (4,-1.8) (xgb) {XGBoost\\(30\%)};
\node[model] at (8,-1.8) (et) {Extra Trees\\(42\%)};

\draw[arr] (train) -- (rf);
\draw[arr] (train) -- (xgb);
\draw[arr] (train) -- (et);

\node[below=0.4cm of rf, font=\scriptsize\itshape] (p1) {Predictions};
\node[below=0.4cm of xgb, font=\scriptsize\itshape] (p2) {Predictions};
\node[below=0.4cm of et, font=\scriptsize\itshape] (p3) {Predictions};

\draw[arr] (rf) -- (p1);
\draw[arr] (xgb) -- (p2);
\draw[arr] (et) -- (p3);

\node[meta] at (4,-4.2) (meta) {Meta-Learner\\(Level 2)};

\draw[arr] (p1) -- (meta);
\draw[arr] (p2) -- (meta);
\draw[arr] (p3) -- (meta);

\node[output] at (4,-5.8) (out) {Final Prediction\\(\texttt{gammaness} / Energy)};
\draw[arr] (meta) -- (out);

\node[left=0.3cm of rf, font=\scriptsize, rotate=90, anchor=south] {Level 1: Base Models};
\end{tikzpicture}%
}
\caption{Architecture of the Stacking Ensemble used for IACT event reconstruction. Training data is fed to three diverse base models (Level~1), each contributing predictions with different strengths. A meta-learner (Level~2) learns the optimal combination of these predictions, producing the final output. The percentages indicate the relative contribution of each base model in the optimized ensemble for gamma/hadron separation~\cite{Pagliaro:2023}.}
\label{fig:stacking}
\end{figure}

\subsection{Case Study: Improving Energy Reconstruction and Reducing Bias}
\label{subsec:5.3}

The power of stacking ensembles has been demonstrated in a case study using simulated data for the ASTRI Mini-Array~\cite{Pagliaro:2023}. For gamma/hadron separation, a stacked ensemble composed of Extra Trees (42\%), XGBoost (30\%), and Random Forest (28\%) provided a notable improvement in the Quality Factor over the baseline Random Forest model, increasing it from 5.94 to 6.74~\cite{Pagliaro:2023}. These figures are computed on large Monte Carlo test samples ($\sim10^5$--$10^6$ events), which provide sufficient statistics to make differences at the second decimal place statistically significant. Note that baseline Quality Factor values reported in the literature for ASTRI Mini-Array analyses depend sensitively on the gammaness threshold and cut-optimization conventions adopted, which can shift the QF by a factor of order two for the same classifier and dataset. The figures quoted throughout this chapter should therefore be interpreted as relative gains over their respective benchmarks rather than as absolute, cross-study performance metrics.

However, the most significant and impactful result was observed in the complementary task of energy reconstruction, for which the meta-learner converges on a different mixture of base models, reflecting the distinct bias structure of that problem. Analysis revealed that models like Random Forest and Extra Trees tended to have a slight positive energy bias (systematically overestimating the energy), while the XGBoost model exhibited a negative bias (systematically underestimating the energy). By creating a Stacking Ensemble composed of 45\% XGBoost, 27.5\% Extra Trees, and 27.5\% Random Forest---weights that are not set by the analyst but are automatically learned by the meta-learner during training as the combination that minimizes the overall loss function~\cite{Pagliaro:2023}---the model was able to effectively balance these opposing biases. The result was a dramatic reduction in the overall energy bias, bringing it very close to zero across a wide and scientifically important energy range from approximately 2 TeV to 100 TeV~\cite{Pagliaro:2023}.

This result marks a qualitative shift in the application of machine learning to this field. The primary goal of early ML applications was to overcome the large statistical uncertainties imposed by the cosmic-ray background. The near-elimination of energy bias addresses a different regime, where systematic rather than statistical uncertainties are the limiting factor---for instance, a residual miscalibration of the energy scale can mimic or hide a spectral cutoff, leading to incorrect inferences about the physics of a source. That a stacking ensemble can self-correct the opposing biases of its constituent models by construction, rather than through ad-hoc post-processing, extends the operational role of machine learning from enabling detection to enabling precision measurement~\cite{Pagliaro:2023}.

\section{Evaluating Performance in High-Energy Astrophysics}
\label{sec:6}

In order to objectively compare different reconstruction methods and to characterize the capabilities of an IACT array, the community has adopted a standardized set of performance metrics. Each metric addresses a specific aspect of the analysis chain: how well the classifier separates signal from background, how accurately the energy and direction of the primary particle are reconstructed, and, ultimately, how faint a source the instrument can detect. We summarize these metrics below.

\subsection{Quantifying Separation Power: The Quality Factor and ROC Analysis}
\label{subsec:6.1}

The primary goal of gamma/hadron separation is to maximize the number of retained gamma rays while minimizing the contamination from the background.
\begin{itemize}
    \item \textbf{Quality Factor (QF):} The most common single metric for evaluating separation power is the Quality Factor, defined as~\cite{Albert:2008gq, Ohm:2009}:
    \begin{equation}
    Q = \frac{\epsilon_{\gamma}}{\sqrt{\epsilon_{bkg}}}
    \label{eq:qf}
    \end{equation}
    where $\epsilon_{\gamma}$ is the gamma-ray acceptance efficiency (the fraction of true gamma rays correctly classified as such) and $\epsilon_{bkg}$ is the background acceptance rate (the fraction of background events misclassified as gamma rays). This metric is a proxy for the statistical significance of a detection, as defined by Li \& Ma~\cite{LiMa:1983}, making it a highly relevant figure of merit.
    \item \textbf{Receiver Operating Characteristic (ROC) Curve:} A more comprehensive view of a classifier's performance is provided by the ROC curve. This curve plots the true positive rate ($\epsilon_{\gamma}$) against the false positive rate ($\epsilon_{bkg}$) for all possible \texttt{gammaness} selection thresholds. The Area Under the Curve (AUC) provides a single-value summary of the classifier's performance, independent of a specific threshold choice~\cite{Pagliaro:2023}.
\end{itemize}

\subsection{Assessing Reconstruction Accuracy: Energy and Angular Resolution}
\label{subsec:6.2}

For the regression tasks, the key metrics quantify the precision and accuracy of the reconstruction. For a set of events, one can build the distribution of $(E_{rec} - E_{true}) / E_{true}$, where $E_{rec}$ is the reconstructed energy and $E_{true}$ is the true (simulated) energy.

\begin{itemize}
    \item \textbf{Energy Resolution:} The energy resolution is defined as the width of this distribution, typically quantified by its standard deviation or the half-width of the interval containing 68\% of the events. It represents the statistical uncertainty on a single energy measurement~\cite{Pagliaro:2023}.
    \item \textbf{Energy Bias:} The energy bias is the mean of the same distribution. A non-zero bias indicates a systematic over- or under-estimation of the energy scale~\cite{Pagliaro:2023}. Minimizing bias is critical for accurate spectral measurements.
    \item \textbf{Angular Resolution:} The angular resolution is defined as the angular separation between the reconstructed and true arrival directions of the primary particles. It is typically quoted as the radius of the circle containing 68\% of the events from a point source and determines the instrument's ability to resolve fine spatial details in astrophysical sources~\cite{Pagliaro:2023}.
\end{itemize}

\subsection{From Classification to Detection: Background Estimation Methods}
\label{subsec:6.3}

The \texttt{gammaness} score produced by the classifier is not used in isolation. To claim a detection, one must estimate the residual number of background events that survive the gamma/hadron separation cut within the signal region around the putative source. This requires a reliable method for estimating the background level. Several standard techniques are used in the field~\cite{deNaurois:2015, Holder:2012}:
\begin{itemize}
    \item \textbf{ON/OFF method:} Dedicated observations of a source-free region of the sky (the OFF region) are used to estimate the background rate. This method is simple but doubles the required observation time.
    \item \textbf{Reflected-region method:} Multiple OFF regions are defined at the same angular distance from the camera center as the ON (source) region, but at different azimuthal positions. Because IACTs have a radially symmetric acceptance to first order, these reflected regions provide a reliable background estimate from the same observation run.
    \item \textbf{Ring-background method:} The background is estimated from an annular region surrounding the source position. This method is particularly useful for generating sky maps of extended sources.
\end{itemize}
The statistical significance of a detection is then computed using the likelihood ratio test of Li \& Ma~\cite{LiMa:1983}, which takes into account the numbers of events in the ON and OFF regions and their relative exposure. The Quality Factor discussed in Section~\ref{subsec:6.1} is directly related to this significance: a higher QF translates into a more significant detection for a given source flux and observation time.

\subsection{High-Level Impact: Instrument Sensitivity}
\label{subsec:6.4}

Ultimately, all the above performance metrics culminate in the single most important characteristic of an IACT: its sensitivity. The differential flux sensitivity is defined as the minimum flux of a point-like source that the instrument can detect with a certain statistical significance (e.g., 5 standard deviations) in a given amount of observation time (e.g., 50 hours), as a function of energy~\cite{Holder:2012, deNaurois:2015}. Better gamma/hadron separation (higher QF), better angular resolution (which reduces the background integration region), and better energy resolution all contribute to improved sensitivity, allowing the instrument to detect fainter sources and characterize them more precisely.

\section{Synthesis and Future Outlook}
\label{sec:7}

This chapter has reviewed the application of machine learning to data reconstruction in Cherenkov telescopes, charting a course from the standard Hillas-parameter-based pipeline to modern frontiers in feature engineering and ensemble modeling. The case studies from the ASTRI Mini-Array demonstrate that significant performance gains can be achieved by innovating along these two axes.

\subsection{Combining Advanced Features and Models}
\label{subsec:7.1}

The key takeaway is that the two frontiers explored—advanced temporal features and advanced ensemble models—are complementary and synergistic. The most powerful analysis pipelines of the near future will likely combine these innovations. A state-of-the-art approach would involve using the full, rich suite of morphological, stereoscopic, and temporal parameters as input features to a sophisticated Stacking Ensemble model. Such a model would leverage the enhanced discriminatory power of timing information while simultaneously benefiting from the bias-reduction and accuracy-boosting capabilities of a stacked architecture. This combined approach represents the current pinnacle of reconstruction techniques that rely on manually engineered features. Table~\ref{tab:summary} consolidates the key performance results discussed in this chapter.

\begin{table}
\caption{Summary of performance improvements achieved by the different analysis strategies discussed in this chapter, using the Quality Factor (QF) for gamma/hadron separation as the primary metric. Results are from Monte Carlo simulations of the ASTRI Mini-Array~\cite{Pagliaro:2023, LaParola:2025}.}
\label{tab:summary}
\begin{tabular}{p{5cm}p{2cm}p{4.5cm}}
\hline\noalign{\smallskip}
Analysis Strategy & QF & Key Improvement  \\
\noalign{\smallskip}\svhline\noalign{\smallskip}
Random Forest (morphological + stereo features) & 3.32 & Baseline \\
RF + temporal features & 3.68 & +10.8\% QF; up to +50\% sensitivity at $E < 1.5$ TeV \\
Stacking Ensemble (morphological + stereo features) & 6.74 & +13.5\% QF vs.\ single RF baseline; near-zero energy bias (2--100 TeV) \\
\noalign{\smallskip}\hline\noalign{\smallskip}
\end{tabular}
\end{table}

\subsection{The Role of Deep Learning: The Next Horizon}
\label{subsec:7.2}

The natural evolution beyond engineered features is to allow the machine learning model to learn the optimal features directly from the raw data. This is the domain of deep learning, which represents the next major horizon for IACT data analysis. Deep learning models---neural networks with many hidden layers---have revolutionized fields such as computer vision and natural language processing over the past decade~\cite{Goodfellow:2016}. Their key advantage is the ability to perform automatic feature extraction: rather than relying on human-designed parameters like the Hillas set, the network learns, through training, which patterns in the raw pixel data are most informative for the task at hand. This paradigm shift has the potential to exploit information that is lost or diluted during the traditional parameterization step. In what follows, we survey the principal deep learning architectures that have been explored for IACT data analysis.

\subsubsection{Convolutional Neural Networks}
\label{subsubsec:7.2.1}

Convolutional Neural Networks (CNNs) are a class of deep neural networks ideally suited for image analysis. The key idea behind a CNN is the use of small learnable filters (also called kernels) that slide across the input image, detecting local patterns such as edges, spots, or gradients. By stacking multiple layers of such filters, the network progressively builds up representations of increasingly complex and abstract features---from simple edges in the first layers to high-level structures (e.g., the overall shape of a shower) in the deeper layers~\cite{Goodfellow:2016}. In the IACT context, instead of calculating Hillas parameters, a CNN can take the 2D camera image of pixel charges directly as input and learn to perform classification and regression~\cite{Shilon:2018, Mangano:2018}. This approach has the potential to capture subtle patterns and correlations in the image that are missed by the Hillas parameterization.

Several CNN architectures have been explored in the field. Shilon et al.~\cite{Shilon:2018} applied deep CNNs based on the well-known ResNet architecture to simulated data from the H.E.S.S. telescope array, demonstrating improvements in gamma/hadron separation over standard Hillas-parameter-based analyses, particularly at low energies where shower images are faint and morphologically ambiguous. Mangano et al.~\cite{Mangano:2018} explored CNNs on simulated CTAO data, showing that the network could learn features that are not easily captured by human-designed parameters. These studies confirm that CNNs can serve as powerful alternatives to the traditional pipeline, especially when the image contains information that is difficult to summarize in a small set of scalar parameters.

Furthermore, the temporal information can be naturally incorporated by treating the map of pixel arrival times as a second input ``channel'' to the CNN, analogous to the color channels in a standard photograph~\cite{Parsons:2020, Spencer:2021}. Spencer et al.~\cite{Spencer:2021} showed that adding the timing channel to the charge image provides a significant boost in background rejection for CTAO, consistent with the findings discussed in Section~\ref{sec:4} for engineered temporal features.

\subsubsection{Recurrent Neural Networks}
\label{subsubsec:7.2.2}

While CNNs treat the camera image as a static 2D snapshot, the Cherenkov signal is intrinsically a time-ordered sequence: pixels are triggered one after the other as the shower front sweeps across the camera. Recurrent Neural Networks (RNNs) are designed precisely for sequential data. An RNN processes its input one element at a time, maintaining an internal ``memory'' (a hidden state) that is updated at each step, allowing the network to learn dependencies across the sequence~\cite{Goodfellow:2016}.

Parsons and Ohm~\cite{Parsons:2020} combined a CNN with a recurrent layer (specifically, a Long Short-Term Memory, or LSTM, unit) to process sequences of camera frames from the H.E.S.S. telescopes. In this architecture, the CNN extracts spatial features from each time frame, and the LSTM learns the temporal evolution of these features across the sequence. This hybrid CNN--RNN approach achieved a reduction of the hadronic background by a factor of two compared to standard analysis techniques at certain energies, providing direct evidence that the temporal development of the shower contains discriminating information that is complementary to the spatial morphology.

\subsubsection{Graph Neural Networks}
\label{subsubsec:7.2.3}

While CNNs operate on regular grids (like images), many real-world data structures are better described as \emph{graphs}---collections of nodes connected by edges. A Graph Neural Network (GNN) is designed to operate on such structures, allowing each node to exchange information with its neighbors and iteratively refine its representation~\cite{Goodfellow:2016}. GNNs are particularly well suited to stereoscopic IACT analysis for two reasons. First, the array of triggered telescopes is naturally described as a graph, where each telescope is a node and the spatial relationships between them (distances, relative orientations) form the edges. Second, unlike CNNs, GNNs can handle a variable number of inputs---a critical advantage for arrays like CTAO, where the number of telescopes participating in an event can vary widely from event to event.

Instead of combining single-telescope reconstructions with a simple weighted average, a GNN learns the optimal way to merge information from multiple viewpoints in a holistic manner. The ``Stereograph'' method, developed for CTAO by Miener et al.~\cite{Miener:2021}, treats each telescope's image-level features as node attributes and lets the network learn how to propagate and aggregate this information across the graph. This approach has shown improved energy and angular resolution compared to standard stereoscopic techniques, suggesting that GNNs may become a key component of future IACT analysis pipelines.

\subsubsection{The Challenge of Domain Shift}
\label{subsubsec:7.2.4}

A key challenge for all supervised learning methods, and particularly for deep learning, is the ``domain shift'' between the simulated data used for training and the real data gathered by the telescopes~\cite{Shilon:2018}. All the models discussed in this chapter---from Random Forests to CNNs---are trained on Monte Carlo simulations in which the properties of the atmosphere, the optical system, and the detector are described by mathematical models. In practice, however, the real observing conditions are never perfectly reproduced: atmospheric transparency varies with weather and aerosol content, mirror reflectivity degrades over time, and the PMT gains fluctuate. These discrepancies mean that the statistical distributions of the input features in real data can differ from those in the simulations, potentially degrading the performance of a model that has been trained exclusively on simulated data~\cite{Holch:2022}.

This problem is particularly acute for deep learning models, which tend to learn very detailed, high-dimensional representations of the input data and can therefore be more sensitive to subtle distributional differences. Several strategies are being explored to mitigate domain shift:
\begin{itemize}
    \item \textbf{Data augmentation:} artificially expanding the training set by applying random perturbations (e.g., adding noise, varying the NSB level) to make the model more robust to real-world variability.
    \item \textbf{Domain adaptation:} training the network to learn features that are simultaneously predictive of the target variable and invariant across the simulation and real-data domains, for example using adversarial training techniques~\cite{Holch:2022}.
    \item \textbf{Transfer learning:} pre-training a model on simulations and then fine-tuning it on a small sample of real data (e.g., observations of a well-known calibration source like the Crab Nebula).
\end{itemize}
Solving the domain-shift problem is arguably the single most important challenge that must be addressed before deep learning methods can be fully deployed in production IACT analysis pipelines.

\subsubsection{Beyond Supervised Learning: Anomaly Detection and Unsupervised Approaches}
\label{subsubsec:7.2.5}

All the methods discussed so far are supervised: they require labeled training data with known ground truth. A complementary and emerging approach is the use of unsupervised or semi-supervised learning, which does not require explicit labels. One particularly promising application is \emph{anomaly detection}. An autoencoder---a neural network trained to compress its input into a low-dimensional latent representation and then reconstruct it---can be trained exclusively on background (hadronic) events. When presented with a gamma-ray event, whose morphology or temporal structure differs from the background, the autoencoder produces a poor reconstruction, yielding a high reconstruction error that can serve as a discriminant variable. This approach has the attractive property of not requiring gamma-ray simulations for training, potentially reducing the dependence on Monte Carlo models and their associated systematic uncertainties.

More broadly, unsupervised techniques such as clustering and dimensionality reduction (e.g., t-SNE, UMAP) can be used to explore the structure of IACT data in a model-independent way, potentially revealing unexpected event populations or instrumental artefacts that would be missed by a supervised classifier trained on a fixed set of categories. While these methods are still in an exploratory phase for IACT analysis, they represent a promising avenue for future research, particularly in the context of the large and diverse datasets expected from CTAO.

\subsection{Conclusion: The Indispensable Role of Machine Learning in the Future of Cherenkov Astronomy}
\label{subsec:7.3}

A final development that will shape the future of machine learning in this field is the move toward open data and reproducible science. CTAO will be the first ground-based gamma-ray observatory to operate as an open observatory, providing calibrated data and high-level analysis tools to the entire astrophysical community~\cite{CTAConsortium:2019}. This open model, supported by standardized software frameworks like ctapipe and Gammapy (Section~\ref{subsec:3.5}), will lower the barrier to entry for researchers from adjacent fields---including the machine learning community itself---and foster the development of novel analysis techniques through open benchmarks and challenges. The availability of common datasets and evaluation criteria will also enable more rigorous and reproducible comparisons between competing methods, accelerating progress in the field.

The evolution of data reconstruction in Cherenkov astronomy is a story of increasing sophistication, driven by the relentless pursuit of greater sensitivity and precision. The journey from simple image-parameter cuts to Random Forests, and now to the frontiers of temporal features, advanced ensembles, and deep learning, reflects the growing complexity and scale of the data. As next-generation observatories like the ASTRI Mini-Array and CTAO come online, they will generate data volumes on the petabyte scale~\cite{Scuderi:2022, CTAConsortium:2019, Vercellone:2022}. With its unprecedented number of telescopes, CTAO will record events at rates far exceeding those of current instruments, and the sheer diversity of its telescope types (LSTs, MSTs, and SSTs) will demand analysis pipelines capable of combining data from cameras with very different characteristics. The challenges are concrete: closing the simulation-to-data domain gap before the first physics runs of CTAO, extending the low-energy threshold below 20\,GeV where transient sources become accessible, and reducing the residual systematic bias on reconstructed spectra to the level required for precision measurements of cutoffs and breaks. Machine learning is the framework in which these challenges will be addressed---no longer an enhancement to a mature pipeline, but the pipeline itself.


\begin{acknowledgement}
The authors thank the editors of this volume for the invitation to contribute this chapter. This work has made use of concepts and results from the ASTRI and CTAO collaborations.
\end{acknowledgement}



\end{document}